\documentclass[pdflatex,sn-mathphys-num]{sn-jnl}

\usepackage{graphicx}%
\usepackage{multirow}%
\usepackage{amsmath,amssymb,amsfonts}%
\usepackage{amsthm}%
\usepackage{mathrsfs}%
\usepackage[title]{appendix}%
\usepackage{xcolor}%
\usepackage{textcomp}%
\usepackage{manyfoot}%
\usepackage{booktabs}%
\usepackage{algorithm}%
\usepackage{algorithmicx}%
\usepackage{algpseudocode}%
\usepackage{listings}%
\usepackage{amsmath}

\begin{document}

\title[Article Title]{Intermolecular Radiative Decay: A non-local decay mechanism providing an insider’s view of the solvation shell}

\author[1]{Johan Söderström}
\email{Johan.Soderstrom@physics.uu.se}

\author[2]{Lucas M. Cornetta}

\author[3]{Victor Ekholm}

\author[4]{Vincenzo Carravetta}

\author[5]{Arnaldo Naves de Brito}

\author[6]{Ricardo Marinho}

\author[1,3]{Marcus Agåker}

\author[3]{Takashi Tokushima}

\author[3]{Conny Såthe}

\author[3]{Anirudha Ghosh}

\author[7]{Dana Blo{\ss}}

\author[7]{Andreas Hans}

\author[8]{Florian Trinter}

\author[9]{Iyas Ismail}

\author[1]{Debora Vasconcelos}

\author[1]{Joel Pinheiro}

\author[10]{Yi-Ping Chang}

\author[10]{Manuel Harder}

\author[11]{Zhong Yin}

\author[1]{Joseph Nordgren}

\author[3]{Gunnar Öhrwall}

\author[1]{Hans Ågren}

\author[1]{Jan-Erik Rubensson}

\author[1]{Olle Björneholm}
\email{Olle.Bjorneholm@physics.uu.se}

\affil[1]{\orgdiv{Department of Physics and Astronomy}, \orgname{Uppsala University}, \city{Uppsala}, \country{Sweden}}

\affil[2]{\orgdiv{Instituto de Física}, \orgname{Universidade de São Paulo}, \city{São Paulo}, \country{Brazil}}

\affil[3]{\orgdiv{MAX IV Laboratory}, \orgname{Lund University}, \city{Lund}, \country{Sweden}}

\affil[4]{\orgdiv{Institute of Chemical and Physical Processes}, \orgname{CNR-IPCF}, \city{Pisa}, \country{Italy}}

\affil[5]{\orgdiv{Institute of Physics Gleb Wataghin}, \orgname{State University of Campinas}, \city{Campinas}, \country{Brazil}}

\affil[6]{\orgdiv{Department of Physics}, \orgname{University of Brasilia}, \city{Brasilia}, \country{Brazil}}

\affil[7]{\orgdiv{Institute of Physics}, \orgname{University of Kassel}, \city{Kassel}, \country{Germany}}

\affil[8]{\orgdiv{Molecular Physics}, \orgname{Fritz-Haber-Institut der Max-Planck-Gesellschaft}, \city{Berlin}, \country{Germany}}

\affil[9]{\orgname{Sorbonne Université, CNRS, Laboratoire de Chimie Physique – Matière et Rayonnement, LCPMR}, \orgaddress{\postcode{F-75005} \city{Paris}, \country{France}}}

\affil[10]{\orgname{European XFEL}, \orgaddress{\postcode{22689} \city{Schenefeld}, \country{Germany}}}

\affil[11]{\orgdiv{International Center for Synchrotron Radiation Innovation Smart}, \orgname{Tohoku University}, \postcode{980-8577} \orgaddress{\city{Sendai}, \country{Japan}}}


\abstract{Aqueous solutions are crucial in chemistry, biology, environmental science, and technology. The chemistry of solutes is influenced by the surrounding solvation shell of water molecules, which have different chemical properties than bulk water due to their different electronic and geometric structure.  It is an experimental challenge to selectively investigate this property-determining electronic and geometric structure. 

Here, we report experimental results on a novel non-local X-ray emission process,  Intermolecular Radiative Decay (IRD),  for the prototypical ions Na$^{+}$ and Mg$^{2+}$ in water. We show that, in  IRD, an electron from the solvation shell fills a core hole in the solute, and the released energy is emitted as an X-ray photon. We analyze the underlying mechanism using theoretical calculations,  and show how IRD will allow us to meet the challenge of chemically selective probing of solvation shells from within.}




\maketitle

\section*{Introduction}

The fundamental importance of water in chemical reactions is intricately linked to the behavior of water molecules in the immediate vicinity of the reactants \cite{atkins2002,fawcett2004,Bunker_2016}.  These solvation shells are interfaces between the solutes and the bulk solvent. Together with a metal ion, they form metal-aqua complexes of fundamental importance in environmental, biological, and applied chemistry. The structure and dynamics in solvation shells differ substantially from bulk water; for example, the water molecules around metal cations are stronger proton donors than the bulk water molecules \cite{Kropman2001,Bunker_2016}. In the case of ionic solutes, the water molecules surrounding an ion are oriented with their positive (negative) end towards the anion (cation), and with increasing ionic charge, the ion-water distance decreases, leading to the ion-water interaction changing from ion-dipole towards dative covalent bonding. The importance of the solvation shell extends beyond small inorganic ions; for example, the functionality of large biomolecules such as proteins is affected by their surrounding hydration shell \cite{Laage2017,Konstantinovsky2022}. 

Several methods are used to determine the molecular geometry of a solvation shell. X-ray and neutron scattering are general methods to measure interatomic distances. They are straightforward to apply to liquids, and they can also provide information about solvation shells \cite{BOWRON20142,Laage2017}. The interatomic distance and coordination around specific solutes can be inferred from measurements of Extended X-ray Absorption Fine Structure (EXAFS) \cite{Filipponi_2001,BOWRON20142}. Nuclear Magnetic Resonance  (NMR) and vibrational spectroscopy selectively probe the structure and dynamics around solutes \cite{fawcett2004,Bakker2008,Laage2017}.  While these techniques successfully provide information about the geometry, investigation of the electronic structure of solvation shells remains an experimental challenge. As the electronic structure is responsible for the chemical interaction of the solute with the neighboring solvent, an experimental tool capable of selective probing would be very valuable. 

X-ray-based spectroscopies are widely used to probe the electronic structure of matter \cite{stohr2023}. Still, progress has been hampered by limited selectivity for solvation shells. The element specificity associated with quasi-atomic core holes is insufficient to separate water molecules in the solvation shell from bulk water. In the soft X-ray range, core-hole decay is typically dominated by non-radiative \emph{local} Auger-Meitner decay, where one valence electron fills the core hole while another is emitted. Non-local decay processes, such as Intermolecular Coulombic Decay (ICD), are sensitive to local neighbors: the decay of a core hole on a solute involves electrons from neighboring water molecules \cite{Jahnke2020}. In fact, such core ICD processes have been observed for ions such as Mg$^{2+}$, Al$^{3+}$, K$^+$, and Ca$^{2+}$ in aqueous solutions \cite{Gunnar10,Wandared11,ICDmanuscript, Mudryk2024}. Very recently, resonant ICD was used to probe the local surroundings of Ca$^{2+}$ in water, including ion pair formation \textbf{  \cite{Dupuy2024}}.
However depth information from these ICD electrons is limited due to the short attenuation length. 
In radiative core-hole decay, X-rays are emitted as outer electrons fill atom-specific local core holes, which allows for more bulk-sensitive studies than using ICD electrons. The radiative decay of solvated ions is largely unexplored. Probing non-local radiative processes of solvated ions to gain information about nearest neighbors, in analogy to non-radiative ICD, has not been reported before for any type of system.

\begin{figure}
    \centering
    \includegraphics[width=1\linewidth]{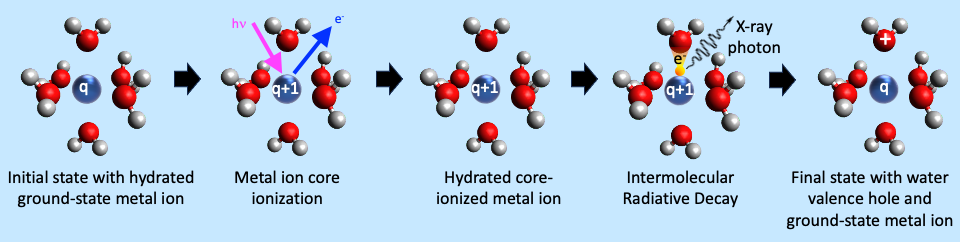}
    \caption{Schematic illustration of the Intermolecular Radiative Decay (IRD) process for a solvated M$^{q}$ ion. From the left, we show the ground state with a M$^{q}$ ion surrounded by six water molecules, followed by core photoionization, which results in a core-ionized M$^{q+1}$ ion.  In the  IRD process, an electron from a solvation-shell water molecule fills the M core hole, and the released energy is emitted as an X-ray photon. In the final state, the M$^{q}$ ion is back to its ground state, and the hole is in the valence levels of the solvation-shell water molecules.}
    \label{fig:IRD_cartoon}
\end{figure}

The electronic configuration of the free metal ions Na$^{+}$ and Mg$^{2+}$ is neon-like,  \textit{1s$^{2}$2s$^{2}$2p$^{6}$}. With the highest occupied states being the \textit{2p} core levels, the most energetic radiative decay of a \textit{1s} core hole is the K$_{\alpha}$ decay, M$^{q+1}$ \textit{1s$^{-1}$} $\rightarrow$ M$^{q+1}$ 2\textit{p$^{-1}$} + h${\nu}$, where M$^{x}$ denotes a metal ion of charge ${x}$. In aqueous solution, Na$^{+}$ and Mg$^{2+}$ ions are surrounded by a solvation shell, which on average consists of six water molecules in an octahedral arrangement at an average ion-oxygen distance of 2.43 Å and 2.07 Å, respectively \cite{Persson2024}. For the solvated Na$^{+}$ and Mg$^{2+}$ ions, any radiative decay at higher photon energy than K$_{\alpha}$ would thus indicate a process involving electrons from the surrounding water molecules. Here, we demonstrate, by measurements and theoretical calculations, how an electron from a neighboring water molecule fills the M \textit{1s} core hole, and the released energy is emitted as an X-ray photon, as schematically shown in Fig.  \ref{fig:IRD_cartoon}. This Intermolecular Radiative Decay (IRD) process enables unique investigations of the solvation shell.

Using the prototypical and well-known Na$^{+}$ and Mg$^{2+}$ ions in water as showcases, we first demonstrate the existence of IRD. Moreover, by theoretical modeling of the process, we show how the IRD depends on factors such as the electronic structure,  ion-water distances, orientational (dis)-order, and chemical composition of the immediate surroundings. This demonstrates the potential of IRD  to probe the solvation shell from within,  providing uniquely selective information on how it differs from bulk water. We foresee that IRD can find applications in studies of the interaction between weakly interacting neighbors, much in the same way as ICD in the non-radiative channel.

\section*{Results and discussion}

\subsection*{IRD: Observation and interpretation}
\begin{figure}[htbp]
   \centering
    \includegraphics[width=0.8\linewidth]{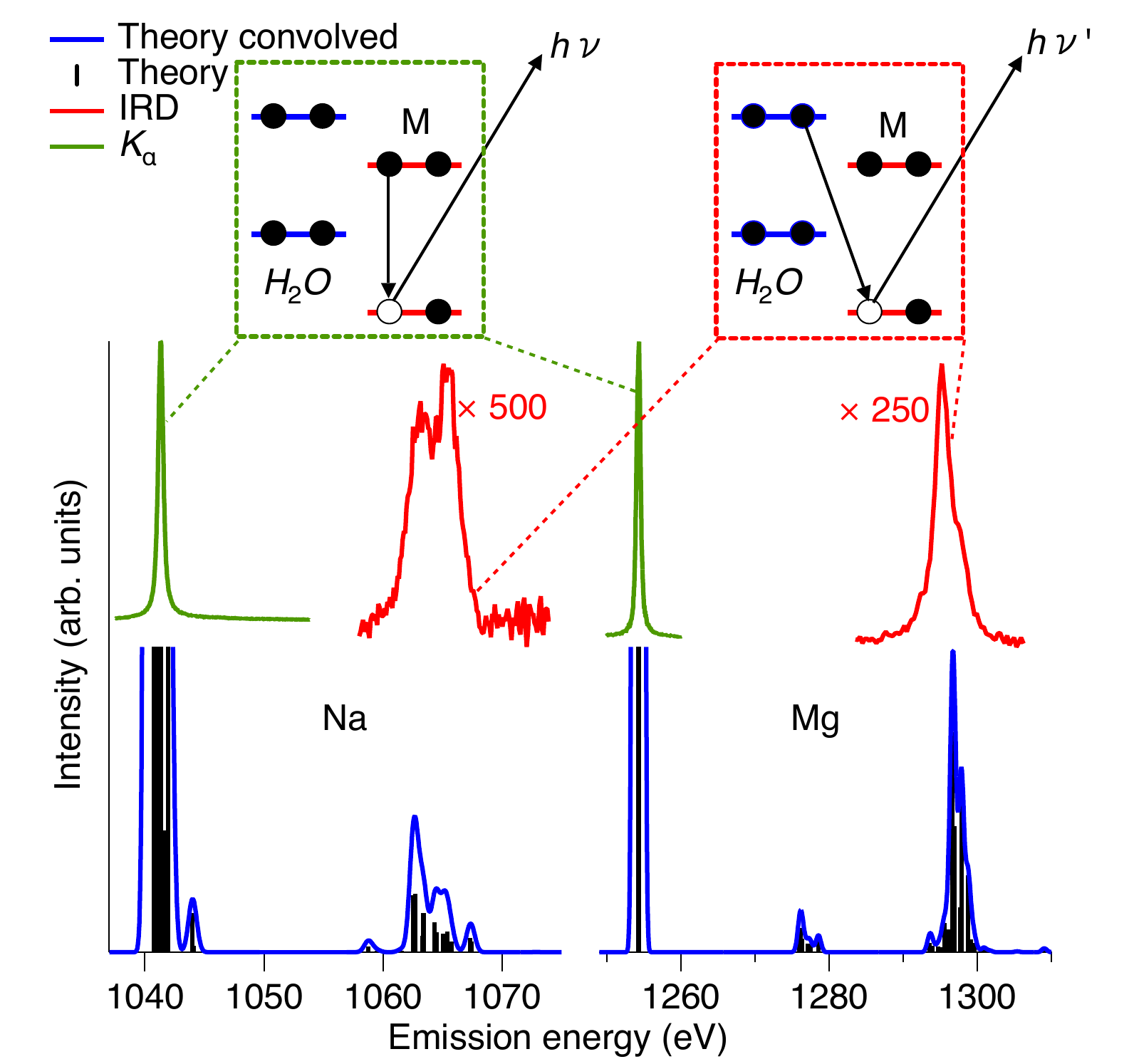}
    \caption{Top: Schematic illustration of the electronic transitions in the local K$_{\alpha}$ decay (green box) and non-local IRD decay (red box). Middle: Experimentally measured X-ray emission spectra for Na$^{+}$ and Mg$^{2+}$ after $1s$ ionization of the metal ion M in the energy range of K$_{\alpha}$ and IRD. For both ions, the spectra exhibit two main features, of which the most intense one is due to the local K$_{\alpha}$ emission, M$^{q+1}$ \textit{1s$^{-1}$} $\rightarrow$ M$^{q+1}$ \textit{2p$^{-1}$} + h${\nu}$. The weaker features at higher photon energies agree well with the energies estimated for the non-local intermolecular radiative decay IRD, M$^{q+1}$ \textit{1s$^{-1}$} + H$_{2}$O $\rightarrow$ M$^{q}$ + H$_{2}$O$^{+}$ \textit{val$^{-1}$}  + h${\nu}$. Bottom: Theoretical X-ray emission spectra for Na$^{+}$ and Mg$^{2+}$ calculated for M$^{q}[\text{H}_2\text{O}]_6$ clusters. The absolute energy scales of the theoretical spectra is shifted so that the  K$_{\alpha}$ emission aligns with the experimental value. The feature around 1277 eV in the Mg spectrum is due to IRD involving the inner valence orbital \textit{2a$_{1}$} of water, which mainly consists of O \textit{2s}. This will not be further discussed, as the corresponding spectral region was not measured.}
    \label{img:spectra}
\end{figure}

Ionization of the $1s$ level of either of the M$^{q}$ metal ions results in an M$^{q+1}$\textit{ 1s$^{-1}$} state, and in Fig. \ref{img:spectra}, we present the X-ray emission spectra from the decay of this core-hole state for solvated Na$^{+}$ and Mg$^{2+}$ ions. The main spectral peak for both ions is located toward the lower photon energy side of the spectra, $\sim$1041 eV for Na and $\sim$1254 eV for Mg. These peaks are due to the K$_{\alpha}$ decay, M$^{q+1}$ \textit{1s$^{-1}$} $\rightarrow$ M$^{q+1}$ \textit{2p$^{-1}$} + h${\nu}$, i.e., a local process, in which the M \textit{1s} hole is filled by a M \textit{2p} electron. 

At somewhat higher photon energy than the K$_{\alpha}$, around $\sim$1060 eV for Na and $\sim$1295 eV for Mg, there are weaker spectral features, which moreover exhibit substructures. As M \textit{2p} is the highest occupied orbital of free Na$^{+}$ and Mg$^{2+}$ ions, these weaker spectral features must involve the neighboring water molecules. 
This can schematically be written as  M$^{q+1}$ \textit{1s$^{-1}$} + H$_{2}$O $\rightarrow$ M$^{q}$ + H$_{2}$O$^{+}$ \textit{val$^{-1}$}  + h${\nu}$, where \textit{val$^{-1}$} denotes a valence hole on the water molecules in the first solvation shell. Just as the K$_{\alpha}$ photon energy is determined by the energy difference between the  M \textit{1s$^{-1}$}  and M  \textit{2p$^{-1}$} states, the photon energy of such non-local radiative decay processes would similarly be given by the energy difference between the M$^{q+1}$ \textit{1s$^{-1}$} + H$_{2}$O and M$^{q}$ + H$_{2}$O$^{+}$ \textit{val$^{-1}$} states. 

In a first approximation, we can use experimentally determined values for M$^{q+1}$ \textit{1s$^{-1}$} from Ref. \cite{ICDmanuscript} and H$_{2}$O$^{+}$ \textit{val$^{-1}$} to roughly estimate the photon energy. For bulk liquid water, the H$_{2}$O$^{+}$ \textit{val$^{-1}$} states span from $\sim$10 to $\sim$20 eV binding energy. Assuming that the H$_{2}$O$^{+}$ \textit{val$^{-1}$} states of the solvation shell are not very different, we use 15 eV as an approximate average value for the H$_{2}$O$^{+}$\textit{ val$^{-1}$} energy.  This yields about 1062 and 1295 eV for the photon energy for such non-local radiative decays for solvated Na$^{+}$ and Mg$^{2+}$ ions, respectively. These values agree quite well with the photon energies of the spectral features observed at higher photon energies, thereby supporting the assignment. 

For an accurate interpretation, X-ray emission spectra for  Na$^{+}$ and Mg$^{2+}$ ions within a cluster of six water molecules have been calculated in the Hartree-Fock (HF) approximation; see the lower panel of Fig. \ref{img:spectra}.  In addition to the local K$_{\alpha}$ decay at lower photon energies, both ions exhibit spectral features at higher photon energies, which coincide very well with the experimentally observed bands. Moreover, we measured the relative intensity of the IRD line vs. the K$_{\alpha}$ line of $\sim$1 $\%$ for Na and $\sim$1.5 $\%$ for Mg. The calculated IRD/K$_{\alpha}$ intensity ratios are 1.1 $\%$  and 1.5 $\%$, respectively, i.e., in excellent agreement with the experimental result. 

Additionally, an analysis of the involved electronic states, to be further discussed below, shows that the contributions to the IRD bands are due to decay from hybrid orbitals mainly composed of water and, to a much lesser extent, metal ion orbitals.

We conclude that the spectral features observed at higher photon energies are due to a non-local radiative decay process, IRD, involving the transfer of an electron from the water molecules in the first solvation shell to the core-ionized metal ion. The IRD process can be seen as a radiative relative of the growing family of non-local decay processes, such as ICD and Electron-Transfer-Mediated Decay (ETMD)  \cite{Jahnke2020,Bloss2024}. 

IRD bears some resemblance to the non-local valence-hole decay process Radiative Charge Transfer (RCT). This process, in which a photon is emitted when a valence electron is transferred from the neutral environment to a charged species with one or more valence holes, is well-known from ion collision studies, see e.g. Refs. \cite{Johnsen1978,Cohen1978}. Closer to the present context, RCT has been indirectly inferred to occur for Ne and Ar dimers \cite{SAITO2007,Kreidi_2008,Yan2013,Higuchi_2010}, and recently directly observed for homogeneous and heterogeneous rare-gas clusters \cite{Hans_2018,Hans2019,Holzapfel2022}. While RCT has been observed in cases where other decay channels are energetically forbidden, the IRD observed here occurs in competition with non-radiative Auger-Meitner decay and local radiative K$_{\alpha}$ decay. Unlike the valence-hole decay process RCT, IRD involves a core-hole, which opens the possibility to use it for chemically selective studies.

We conclude that the calculated spectra agree quite well with the experiments in terms of energy positions of the IRD features and their intensity relative to \textit{K$_{\alpha}$}. This gives us confidence to use the calculations in combination with the experimental results to further explore the mechanisms of the IRD decay, how it depends on the surroundings of the ion, and to discuss how IRD can provide an insider's view of the first solvation shell.

\subsection*{The IRD mechanism }

To investigate the mechanisms behind IRD, specifically what gives intensity and where the electron filling the core hole comes from, we will start by examining a simplified type of model system, M$^{q}[\text{H}_2\text{O}]_6$. This consists of a metal cation M$^{q}$ surrounded by six water molecules in a symmetric $D_{2h}$ structure, see Fig.   \ref{fig:M_H2O_6_model_structure} in the SI. In the one-particle picture, the electronic structures of these systems can be described by orbitals, here denoted hybrid orbitals ${\psi}_{\mu}$, each of which is built by a Linear Combinations of Atomic Orbitals (LCAO) ${\chi}_{\nu}$. As usual, the atomic orbitals are a Cartesian-Gaussian functions set centered on the metal ion and on all water-related atoms. For simplicity, the $\chi$ has been indexed simply by $\nu$, in such a way that one can write ${\psi}_{\mu} = \sum_\nu{C}_{\mu}^\nu{\chi}_{\nu}$, where $C_\mu^\nu$ is the hybrid orbital coefficient. The most relevant hybrid orbitals ${\psi}_{\mu}$ of the symmetric model are shown in Fig. \ref{fig:na_panel_orb} of the SI. 

\begin{figure}
    \centering
    \includegraphics[width=0.8\linewidth]{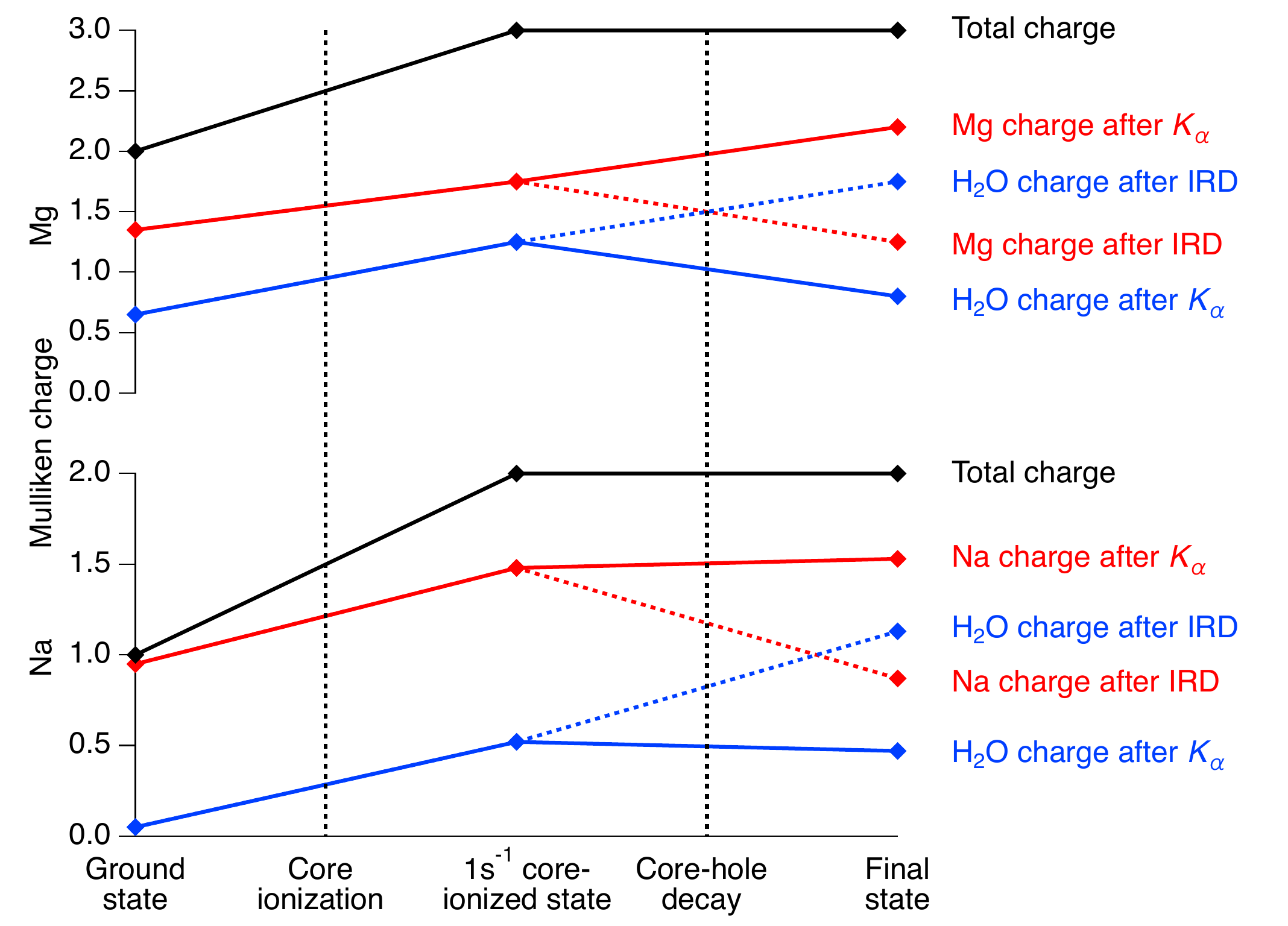}
    \caption{Mulliken charges in the ground, intermediate, and final states for Na$^{+}$[H$_{2}$O]$_{6}$  (lower panel) and Mg$^{2+}$[H$_{2}$O]$_{6}$  (upper panel). The total charge (black) is decomposed into charge on the metal ion (red), and on the water molecules (blue). For both metal ion and water, the final-state charge after both local K$_{\alpha}$ decay and non-local IRD is shown. The connecting lines are only guides for the eye.}
    \label{fig:Mulliken}
\end{figure}

As a starting point, we will follow the development of the charge distribution from the initial state, over the core-ionized state, to the different final states reached by radiative decay. Figure\ref{fig:Mulliken} shows the Mulliken charges of the M$^{q}$[H$_{2}$O]$_{6}$ ions in the three states, see also Table \ref{tab:Mulliken_table} in the SI. For a free, isolated Na/Mg ion, the initial state has a charge of +1/+2, and both the core-ionized and the final states have a charge of +2/+3, as indicated in black. For the solvated ions, the total charge develops in the same way, but the presence of the water molecules allows for charge redistribution. In the ground state of the Na$^{+}$[H$_{2}$O]$_{6}$ ion cluster, the Na ion charge is +0.95, i.e., very close to +1. A small charge of +0.05 is delocalized onto the six water molecules, which means that the Na$^{+}$-water interaction is quite close to an ideal ion-dipole interaction. For the Mg$^{2+}$[H$_{2}$O]$_{6}$ ion cluster, the charge on the Mg ion is +1.35, with +0.65 on the six water molecules. The Mg$^{2+}$-water interaction does thus contain charge transfer via hybridization leading to some dative covalent bonding in addition to the ion-dipole interaction.

Core ionization increases the total charge to +2 for the Na case and to +3 for the Mg case. In the core-ionized Na$^{2+}$[H$_{2}$O]$_{6}$ ion cluster, the Na ion has charge +1.48 instead of +2, implying +0.52 on the waters. In contrast to the ground state, there is now a charge transfer from the water to the ion, amounting to 0.52 of an electron. For the core-ionized Mg$^{3+}$[H$_{2}$O]$_{6}$ ion, the +3 total charge is even more distributed, with +1.75 on Mg and +1.25 on the water. 

In the present context, the charge transfer upon core ionization and core-hole decay can be regarded as instantaneous, but the ultrafast dynamics of this may be possible to probe in the not-too-distant future, using e.g., attosecond X-ray pulses from free-electron lasers. 

The radiative core-hole decay does not change the total charge, but the charge distribution is entirely different after local  K$_{\alpha}$ decay and IRD. For Na and Mg,  K$_{\alpha}$ decay \textit{increases} the charge of the metal ion and \textit{decreases} the charge on water, compared to the intermediate core-hole state with equal total charge. In contrast, IRD instead \textit{decreases} the charge of the metal ion and \textit{increases} the charge on water. In the intermediate core-hole state, most of the charge is on the metal ion. The final state after  \textit{K}$_{\alpha}$ decay also has the majority of the charge on the metal ion, but after IRD, the majority of the charge is instead on the water. This clearly shows that the net effect of the IRD process is that the charge from the water fills the core hole on the metal ion.

We can take the discussion of the IRD process one step further by analyzing what mechanism provides the intensity of the IRD transitions. A detailed analysis using the LCAO model, see the SI, yields that for IRD, the intensity and the electron come from different sources. Even though the hybrid orbitals have relatively small metal character, the associated high dipole transition moments of these metal-centered orbitals result in a much higher contribution to the intensity than from the oxygen-centered orbitals. At the same time, the final-state hole is delocalized on the nearest water molecules, meaning that the electron filling the M \textit{1s} core hole effectively mainly comes from the solvation shell water. In spite of  this non-local character of IRD, the strong contribution to the intensity from the small metal character of the hybrid orbitals makes the one-center approximation remain a good model.
Moreover, by varying the M-O distance, the IRD efficiency is seen to decrease rapidly with increasing distance, as discussed in the SI. This means that IRD to a good approximation only involves the first solvation shell.

\subsection*{Electronic structure}
\begin{figure}
    \centering
    \includegraphics[width=0.8\linewidth]{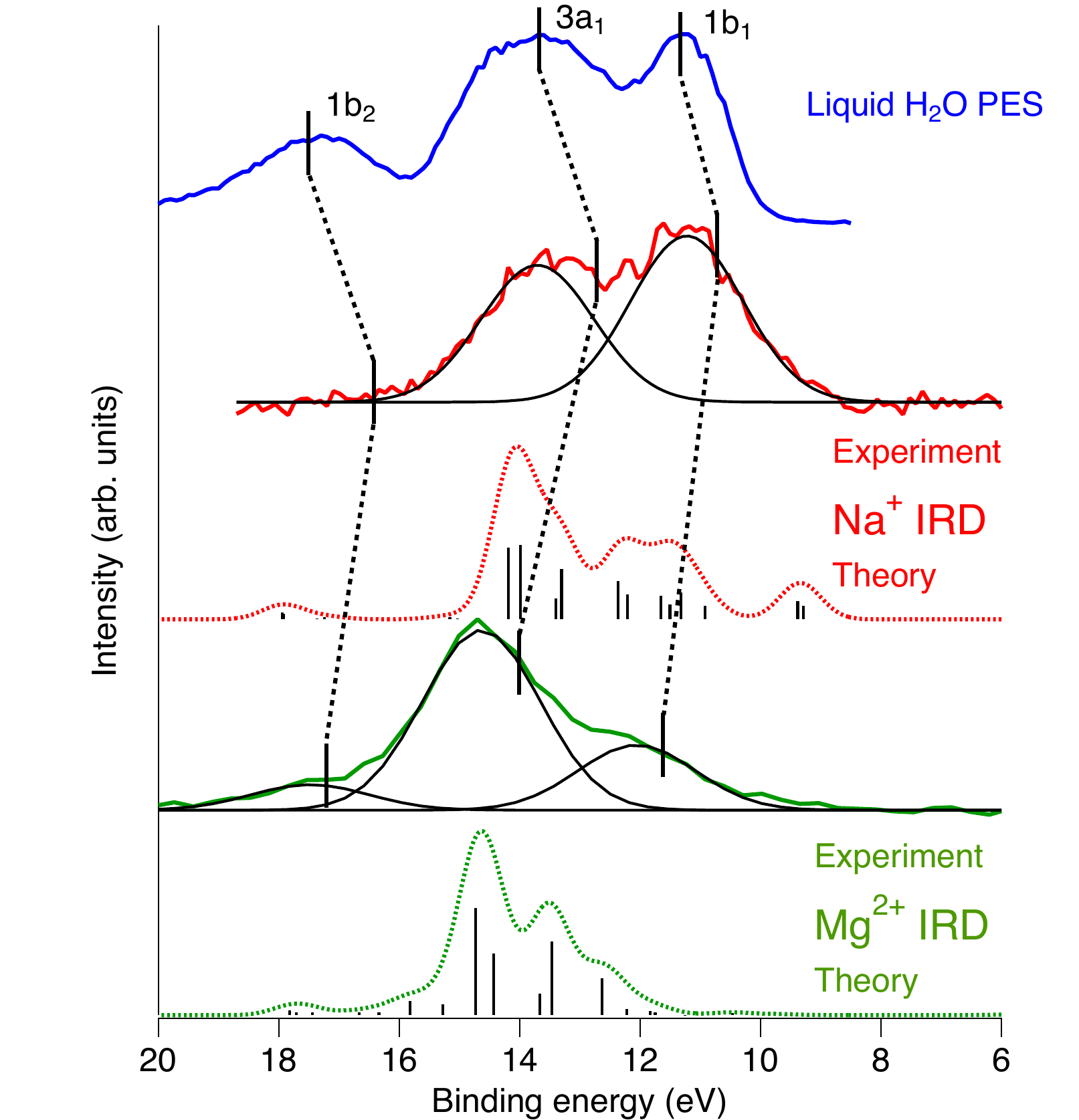}
    \caption{Comparison of one-hole spectra. From the top, liquid-water PES spectrum recorded using a photon energy of 532 eV, i.e. under relatively bulk-sensitive conditions \cite{Thurmer_PRL_2013} (blue), Na IRD spectrum (red, experimental solid and  theory dashed), Mg IRD spectrum (green, experimental solid and theory dashed). Experimental and theoretical IRD spectra are enlargements of the data shown in Fig.  \ref{img:spectra}. Solid black curves indicate a fitting of the experimental data using a minimal number of Gaussian curves. Original theoretical data for Na and Mg are represented by black bars, and the dashed curves for their convolution with a Gaussian function (FWHM = 0.8 eV) to facilitate the comparison with the experimental spectra. Vertical solid lines indicate calculated binding energies from Ref. \cite{ICDmanuscript}.}
    \label{fig:valence bands}
\end{figure}

As we will show, IRD offers a chemically selective way to probe the electronic structure of the first solvation shell around a solvated ion. Starting with bulk liquid water, the electronic structure has been studied by valence band photoelectron spectroscopy (PES), providing the H$_{2}$O$^{+}$ \textit{val$^{-1}$} energies, see e.g. Ref. \cite{Thuermer2021}. The valence band photoelectron spectrum of liquid water, shown in Fig.  \ref{fig:valence bands}, consists of three features. These are associated to ionization of the three outermost valence orbitals of the water molecule, \textit{1b$_{1}$},\textit{ 3a$_{1}$}, and \textit{1b$_{2}$}, modified by overlap and screening in the liquid environment \cite{Winter04}.  
As a simplistic first approximation, the solvation-shell H$_{2}$O$^{+}_{solv}$ \textit{val$^{-1}$} binding energies would be expected to increase around a positively charged ion, but we should also consider that the geometrical arrangement of the close to six water molecules surrounding a cation is quite different from that of the tetrahedral-like structure of the close to four water molecules surrounding a typical water molecule in bulk liquid water. The water-water orbital interactions can thus be expected to be different in the solvation shell, and in addition there may be ion-water orbital interactions, including polarization of the orbitals by the electric field of the ion. We can obtain the corresponding one-hole energies for the solvation-shell water, E(H$_{2}$O$^{+}_{solv}$ \textit{val$^{-1}$}), from the IRD spectra as the difference between the M \textit{1s} binding energy of the metal ion, E(M \textit{1s$^{-1}$}),  and the photon energy of the IRD transitions, h$\nu$(IRD); E(H$_{2}$O$^{+}_{solv}$ \textit{val$^{-1}$}) = E(M \textit{1s$^{-1}$}) - h$\nu$(IRD).  

Turning to the Na IRD spectrum, shown in Fig.  \ref{fig:valence bands}, it is seen to exhibit two peaks with energies very close to the two most intense peaks of bulk water. As for bulk water, we provisionally interpret these as corresponding to holes in the \textit{ 3a$_{1}$} and \textit{1b$_{1}$} orbitals in the first solvation shell.  For Mg, there are three peaks, of which the two stronger ones exhibit clear shifts to higher energies relative to bulk water.  

Figure \ref{fig:valence bands} also shows calculated IRD spectra, obtained by HF calculations for Na$^{+}$[H$_{2}$O]$_{6}$ and Mg$^{2+}$[H$_{2}$O]$_{6}$ in a representative cluster geometry derived from molecular dynamics simulations, see Fig. \ref{fig:M_H2O_6_model_structure}.  For both ions, the calculations and the experiments agree quite well  in terms of energies. For the spectral shape, there is also a good agreement for Mg with\textit{ 3a$_{1}$} dominating over \textit{1b$_{1}$} and\textit{ 1b$_{2}$}. For Na, the calculated spectrum also peaks in the\textit{ 3a$_{1}$} range, whereas for the experimental spectrum \textit{1b$_{1}$} is even stronger than \textit{3a$_{1}$}. We tentatively ascribe this discrepancy as due to orientational disorder of the water molecules in the solvation shell, see the SI.. 

It should be noted, however, that the calculations show that the experimentally measured peaks cannot uniquely be assigned to holes in the \textit{1b$_{1}$}, \textit{3a$_{1}$}, and \textit{ 1b$_{2}$} molecular orbitals of individual solvation-shell molecules. Instead, the final one-hole states are delocalized over two or more solvation-shell molecules. The orbital composition and energy of the final one-hole states also vary with the different relative orientations of the water molecules. This means that spectral contributions from holes in the three valence molecular orbitals of water overlap.  That said, the relative contribution of \textit{1b$_{1}$} holes is highest for the low-energy states, whereas the relative contribution of \textit{3a$_{1}$} holes is highest for states of higher energy. We will, therefore, with the above reservations in mind, continue to denote the IRD spectral features as \textit{1b$_{1}$}, \textit{3a$_{1}$}, and \textit{ 1b$_{2}$}.

We can also compare the observed H$_{2}$O$^{+}_{solv}$ \textit{val$^{-1}$} energies for Na and Mg with calculated energies from Ref. \cite{ICDmanuscript}, shown as dashed vertical lines in Fig. \ref{fig:valence bands}. Contrary to the simplistic expectation, the calculations predict a small shift towards lower one-hole energies relative to pure water for Na. This was explained as due to the structure-breaking properties of Na, causing a disruption of the water hydrogen-bond network. 

For Mg, our results and the calculations from Ref.  \cite{ICDmanuscript}  agree upon a shift towards higher energy relative to pure water. For both ions, the calculated energies from Ref. \cite{ICDmanuscript} are somewhat lower than our observed ones. This may be due to the experimentally obtained H$_{2}$O$^{+}_{solv}$\textit{ val$^{-1}$} states being vibrationally excited, whereas the calculated energies are for the vibrational ground states.

The present results based on IRD can also be discussed relative to the O \textit{1s} XES spectra of solvation-shell water obtained by subtracting a suitably scaled spectrum for pure water from a spectrum of a MgCl$_{2}$ solution \cite{Yin2017}. The two approaches probe the same electronic final states, H$_{2}$O$^{+}_{solv}$ \textit{val$^{-1}$}, but in different ways. The spectra obtained by the subtractive procedure in Ref.  \cite{Yin2017} contain contributions from solvation shells around both Mg$^{2+}$ and Cl$^{-}$ ions, whereas IRD probes the solvation shells around the different ions selectively. The spectral shapes are also quite different, as the processes are very different, M$^{q+1}$ \textit{1s}$^{-1}$  + H$_{2}$O$_{solv}$ $\rightarrow$ M$^{q}$ + H$_{2}$O$^{+}_{solv}$ \textit{val$^{-1}$} + h${\nu}$  for the present IRD spectra, versus H$_{2}$O$^{+}_{solv}$ O \textit{1s}$^{-1}$  $\rightarrow$  H$_{2}$O$^{+}_{solv}$ \textit{val$^{-1}$}  + h${\nu}$ for the differential O \textit{1s} XES spectra of Ref. \cite{Yin2017}. This results in different relative intensities of H$_{2}$O$^{+}_{solv}$ \textit{val$^{-1}$ }electronic states, as well as different populations of vibrational states. The spectra obtained by the subtractive procedure are also influenced by the nuclear dynamics during the O \textit{1s} core-hole lifetime, a complication not affecting the IRD spectra.

We conclude that IRD can be used to selectively probe the  H$_{2}$O$^{+}_{solv}$ \textit{val$^{-1}$} energies of the first solvation shell around ions, revealing how the electronic structure differs from bulk water. This would also include the presence of other species, e.g., contact ion pairs, see the SI.

\section*{Summary and conclusions}
Combining experiments and calculations, we have demonstrated the existence of a novel core-hole decay process, Intermolecular Radiative Decay (IRD), for Na$^{+}$ and Mg$^{2+}$ ions in aqueous solution. In IRD, a $1s$ core hole on Na or Mg is filled by an electron from a neighboring species, and an X-ray photon is emitted. IRD is thus a non-local decay process, which in some sense can be considered a radiative analog of Interatomic Coulombic Decay (ICD). 

 The transitions observed in the IRD spectra are shown to get intensity from hybridization between valence orbitals on water molecules and the occupied orbitals on the metal cation. 
The associated high atomic transition moments result in a much higher contribution to the intensity than from the dominant oxygen character. At the same time, the final-state hole is delocalized on the nearest water molecules, meaning that the electron filling the metal \textit{1s} core hole effectively  comes from water. For the observed IRD lines, the intensity comes from the electron density on the metal ion, while the energy position derives from the electron density delocalized on the water molecules.

The IRD spectra are shown to reflect fundamental properties of the solvation shell, e.g., its radius, composition including possible ion pairing, electronic structure, and orientational disorder.

These results for hydrated Na$^{+}$ and Mg$^{2+}$ ions demonstrate the potential of IRD as a more generally applicable method to selectively probe the solvation shell from within, thereby opening new possibilities to understand how the solvation shell affects the chemical properties of solutes.

\section*{Methods}

\subsection*{Experimental}
The measurements for Na$^{+}$ and Mg$^{2+}$ in water were performed using a cylindrical liquid jet at the VERITAS  beamline at the MAX IV synchrotron radiation facility in Lund, Sweden \cite{maxiv_veritas}. The VERITAS beamline comprises an elliptically polarizing undulator and a collimated plane-grating monochromator with an ellipsoidal refocusing mirror. The beamline is equipped with a large constant-line-spacing grating Rowland spectrometer with a collimating mirror in the non-dispersive direction and a cylindrical grating with 1400 grooves/mm and 67 m radius.  Measurements were done with linear horizontal polarization of the incident radiation. 

The samples, 2 M NaCl and 2 M MgCl$_{2}$ solution, were prepared by dissolving commercially purchased NaCl and MgCl$_{2}$ (Sigma-Aldrich) with purity  of $>98\%$  in MilliQ (18.2~M${\Omega}$/cm) water. The sample was introduced via a vertically mounted liquid jet shooting into a cold trap cooled by liquid nitrogen. The liquid jet was surrounded by a cylindrical differential pumping stage with holes for incident and outgoing photons, allowing the sample to be intersected by the X-rays from the beamline in front of the soft X-ray spectrometer. 

Ionization of the \textit{1s} level of the Na$^{+}$ and Mg$^{2+}$ ions was done using photon energies of 1083 and 1318 eV, respectively, i.e., 6 eV and 8 eV above the respective M \textit{1s} thresholds. In both cases, these photon energies are well below the thresholds for the formation of higher-energy states, like M \textit{1s$^{-1}$2p$^{-1}$3p} or M\textit{ 1s$^{-1}$2p$^{-1}$}, ensuring that M \textit{1s$^{-1}$} is the only state from which the decay occurs.
Emitted photons were detected in the plane of the polarization of the incident radiation at an angle of $90^{\circ}$ with respect to the propagation axis. Calibration of the emitted photon energies was done using the energy of the K$_{\alpha}$ transition, M\textit{ 1s$^{-1}$} $\rightarrow$ M  \textit{2p$^{-1}$} +  h${\nu}$, to set the absolute energy at one point of the detector, and a set of elastic peaks to determine the dispersion over the detector. The energy of the K$_{\alpha}$ transition was determined from the binding energy difference between the M \textit{1s$^{-1}$} and M\textit{ 2p$^{-1}$} states of solvated  Na$^{+}$ and Mg$^{2+}$ ions, respectively, obtained by electron spectroscopy \cite{ICDmanuscript}. For  Na$^{+}$ and Mg$^{2+}$, we thus obtain h${\nu}$(\textit{K$_{\alpha}$}) as 1041.3 eV  (1076.7 - 35.4) and 1254.3 eV (1309.9 - 55.6), respectively  \cite{ICDmanuscript}. This procedure relies on the energy difference between the elastic peaks and not their absolute energy, thereby becoming independent of the absolute monochromator energy scale. 
An upper limit of the spectrometer resolution is the total observed width of the K$_{\alpha}$ peaks. For both the Na$^{+}$ and Mg$^{2+}$ edges, this is less than 0.6 eV, which is significantly smaller than the width of the fitted components of the IRD signals ($\sim$2.2 eV). The X-ray emission spectra of Na  and Mg were measured in the third and second order of diffraction, respectively.

A pilot study for Mg$^{2+}$ in water was also performed at the P04 beamline of the synchrotron radiation facility PETRA III (DESY, Hamburg, Germany), see SI. The result from P04 is consistent with the results from VERITAS, on which the analysis in this paper is based.

The valence band photoelectron spectrum of water was recorded at the FlexPES beamline of MAX IV, Lund, Sweden \cite{Flexpes}, using a photon energy of 532 eV. The photon energy resolution was $\sim$0.23 eV, and the spectrometer resolution $\sim$0.25 eV, resulting in a total resolution of $\sim$0.35 eV. To reduce spectral contributions from the gas phase water, the spectrum was recorded using a bias of 50 V.

\subsection*{Theory}
Theoretical investigations were conducted to analyze the experimental spectrum and to understand the mechanisms behind the IRD effect. \textit{Ab initio} and Density-Functional Theory (DFT) calculations of the XES spectrum for two kinds of models were carried out: small idealized systems and, respectively,  ion/water clusters inferred by molecular dynamics simulations of aqueous solutions of NaCl and MgCl$_2$. Using the small systems, as M$^{q}$[H$_2$O]$_n$, with $n=1,4,6$, the role of distance, symmetry, and number of molecules surrounding the cation were studied. Both models were used to compare simulated and observed experimental spectra, including the intensity ratio between K$_\alpha$ and IRD transitions, and to rationalize the mechanisms behind the IRD process.  A one-center approximation is formulated for IRD and is evaluated with full calculations in order to verify the validity of this approximation.

\subsubsection*{Small models}
Due to the positive charge of the M$^{q}$ metal cation, the distribution of the closest surrounding water molecules is expected to point with the partly negatively charged oxygen atoms inwards toward the metal ion; our small models are based on such an assumption. The water molecules were oriented with the dipole vector along the  M$^{q}$--O axis ($R$), in a planar geometry for M$^{q}$[H$_2$O]$_4$, and in an octahedral geometry for M$^{q}$[H$_2$O]$_6$. 

Transition energies and radiative decay rates were studied within the RASSCF level of theory, as implemented in the OpenMolcas package \cite{openmolcas}, together with the ANO-RCC-VQZP basis set and using the Hartree-Fock wavefunction as reference (HF/RAS). 
For the different M$^{q}$[H$_2$O]$_n$ cases, the RAS subspaces were built with the M$^{q}$ \textit{1s} orbital in RAS1, the three M$^{q}$ \textit{2p} orbitals in RAS2, and the 4$n$ water molecular orbitals corresponding to \textit{2a$_{1}$}, \textit{1b$_{2}$}, \textit{3a$_{1}$}, and \textit{1b$_{1}$} of the $n$ water molecules in RAS3, in such a way that the ionized system had a total of 2(4 + 4$n$)-1 active electrons in 4 + 4$n$ orbitals. Relativistic effects based on the Douglas-Kroll decomposition of the electronic Hamiltonian and spin-orbit couplings were included in both calculations.

\subsubsection*{Cluster models}
To determine a more realistic structural model of Na$^+$ or Mg$^{2+}$ ions solvated in water, molecular dynamics calculations were carried out with the GROMACS code version 5.1.4 \cite{abraham_2015} on two samples, both contained in a cubic box with a side length of 50 \textup{\AA} and periodicity conditions in the three dimensions. The first sample, which was intended to simulate a situation of extreme dilution, contained a single M$^{q}$ ion and the appropriate number of Cl$^{-}$ counter ions, and 4166 water molecules, while in the second case, the number of M$^{q}$ ions was brought to 150 and the counter ions increased correspondingly, to replicate the concentration of the solution (2 M) used experimentally. 
For the M$^{q}$ and Cl$^{-}$ ions, the force field OPLS/AA \cite{jorgensen_1996} has been used, while the water molecule has been simulated with the TIP4P model \cite{jorgensen_1983}. Starting from a random distribution of the molecules in the solution, a standard minimization-heating-equilibration protocol was adopted, followed by an extended production run of 10 ns with a time step of 1 fs. Apart from the heating-equilibration steps, the canonical ensemble (NVT) dynamics were conducted at T = 300 K fixed with a velocity-rescaling thermostat. 

The analysis of the molecular dynamics simulation carried out on the first sample confirms the result of accurate quantum molecular dynamics calculation \cite{ditommaso_2010}, i.e., that, at high dilution, the M$^{q}$ ion is surrounded by six water molecules with high coordination. In contrast, the second hydration shell is less ordered and more sensitive to rearrangements at room temperature.
From the analysis of the molecular dynamics trajectories, the most typical structures of the solvation shells of M$^{q}$ in a very diluted aqueous solution and at the 2 M concentration were identified. 
This allowed us to select some specific clusters containing a single M$^{q}$ ion and a variable number of water molecules for which the XES spectrum generated by the ionization of the M$^{q}$ core shell was calculated. This was accomplished in the Kohn-Sham DFT ground-state approximation utilizing the StoBe-deMon code
\cite{stobe_2014} the Perdew-86 density functional, and a TZVP basis set \cite{godbout_1992}. Such a method can be easily applied to large systems, but is affected by the well-known self-interaction error for the core electron. The HF approximation, fully including the electron relaxation around the initial core hole, can be considered, despite missing any electron correlation effect, to provide a more reliable simulation of the XES spectra. 
However, the present calculations found that both DFT methods, with an appropriate energy shift, and HF methods predict quite similar XES spectra.

\backmatter

\bmhead{Supplementary information}
Supplementary information available.

\bmhead{Data availability}
Data to recreate the figures in this paper is available in digital format in Ref. \cite{this_dataset}.

\bmhead{Acknowledgements}
We acknowledge MAX IV Laboratory for time on the VERITAS beamline under proposal 2022-1207. Research conducted at MAX IV, a Swedish national user facility, is supported by the Swedish Research Council under contract 2018-07152, the Swedish Governmental Agency for Innovation Systems under contract 2018-04969, and FORMAS under contract 2019-02496. We acknowledge DESY (Hamburg, Germany), a member of the Helmholtz Association HGF, for the provision of experimental facilities. Parts of this research were carried out at PETRA III and would like to thank Moritz Hoesch and his team for assistance in using beamline P04. Beamtime was allocated for proposal I-20211465 EC. O.B.,  J.-E.R. and H.{\AA}. acknowledge funding from the Swedish Research Council (VR) through the projects VR 2023-04346, VR 2021-04017, and VR 2022-03405, respectively, as well as the Swedish Foundation for International Cooperation in Research and Higher Education (STINT) through project 202100-2932. The authors also thank the Swedish National Infrastructure for Computing (SNIC 2022/3-34) at the National Supercomputer Centre of Linköping University  (Sweden) partially funded by the Swedish Research Council through grant agreement no. 2018-05973. M.H. acknowledges funding by the German Federal Ministry of Education and Research under grant number 13K22XXB DYLIXUT. D.B. and A.H. acknowledge support by the German Federal Ministry of Education and Research (BMBF) through project 05K22RK1-TRANSALP. F.T. acknowledges funding by the Deutsche Forschungsgemeinschaft (DFG, German Research Foundation) - Project No. 509471550, Emmy Noether Programme.

\bibliography{Draft_main_V1}

\newpage


{\centering\Large\bfseries Supplementary Information for: Intermolecular Radiative Decay: A non-local decay mechanism providing an insider’s view of the solvation shell\par}

\tableofcontents

\section{Experimental Results from P04 at PETRA III}
\begin{figure}[!htbp]
   \centering
    \includegraphics[width=0.8\textwidth]{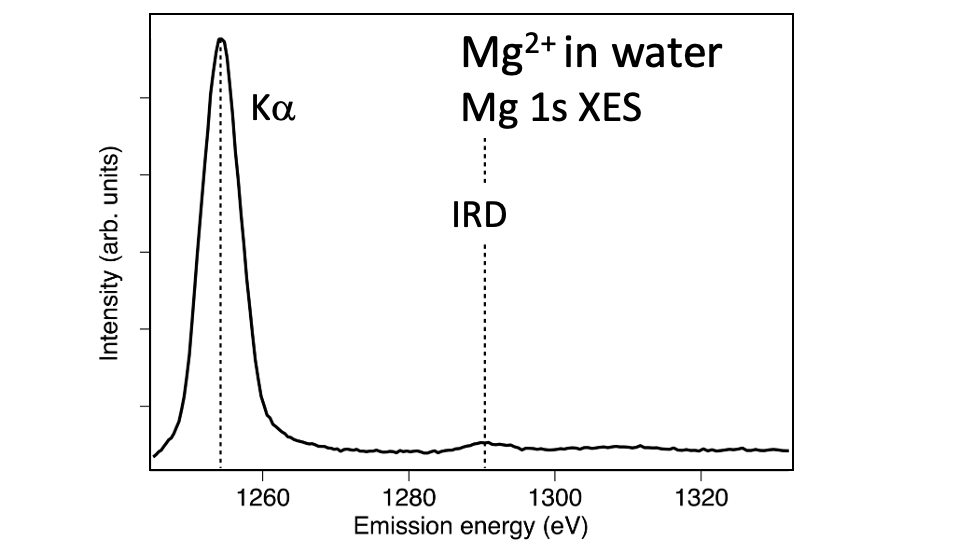}
    \caption{The experimental  X-ray spectrum after Mg$^{2+}$ $1s$ ionization in the energy range of K$_{\alpha}$ and IRD measured at P04. The spectrum shown is a sum of several spectra recorded with different photon energies above and just below the Mg$^{2+}$ $1s$ ionization threshold. Within the experimental resolution these spectra exhibit no significant differences. To improve the statistics, the spectra were therefore summed. The spectrum contains two peaks, of which the most intense one is the K$_{\alpha}$, Mg$^{3+}$\textit{1s}$^{-1}$ $\rightarrow$ Mg$^{3+}$\textit{1s}$^{2}$\textit{2p}$^{-1}$ + h$\nu$. The weaker peak at 1290.4 eV agrees well with the energy estimated for the non-local radiative decay, Mg$^{3+}$\textit{1s}$^{-1}$ + H$_{2}$O $\rightarrow$ Mg$^{2+}$ + H$_{2}$O$^{+}$ val$^{-1}$  + h$\nu'$. The vertical dashed lines are guides for the eye for the K$_{\alpha}$ and IRD features.}
    \label{img:spectra2}
\end{figure}

For Mg$^{2+}$ in water, preliminary experiments were also carried out at the P04 beamline of the synchrotron radiation facility PETRA III, DESY, Hamburg  \cite{viefhaus2013variable}, using a setup combining a cylindrical liquid jet and a soft X-ray spectrometer  \cite{Nordgren1989}, see Fig. \ref{img:spectra2}. The sample, 2M MgCl$_{2}$ solution, was prepared by dissolving commercially purchased MgCl$_{2}$ (Sigma-Aldrich with purity  of $>98\%$) in MilliQ (18.2~M${\Omega}$/cm) water. The sample was introduced via a vertically mounted liquid jet shooting into a liquid nitrogen cooled cold trap. The liquid jet was intersected by the circularly polarized X-rays from the beamline in front of the soft X-ray spectrometer. The spectrometer was a modified grazing-incidence Rowland-circle spherical grating spectrometer (GRACE) equipped with a 1200 l/mm grating \cite{Nordgren1989}. The X-ray emission spectra were measured in fourth order.  The beamline and the X-ray detector were protected from the relatively high pressure in the measurement chamber during liquid jet operation by a differential pumping stage and a filter between the grating and the detector, respectively. Due to the presence of the filter, the internal motions in the spectrometer were impeded. The detector could thus not be positioned in the optimal focus, resulting in a resolution of $\sim$ 6 eV.

\section{Additional Theoretical Results}
\subsection{Model system structures}
\begin{figure}[!htbp]
    \centering

    \includegraphics[width=0.3\linewidth]{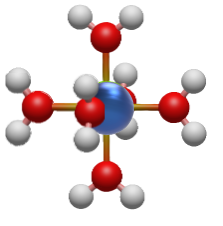}
    \newline
    Symmetric M$^{q}$(H$_2$O)$_6$ structure
    
    \includegraphics[width=0.3\linewidth]{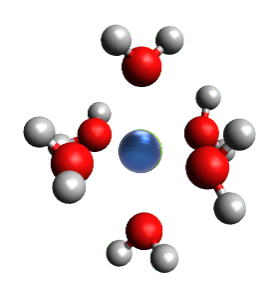}
    \newline
    Representative distorted M$^{q}$(H$_2$O)$_6$ structure

    \includegraphics[width=0.3\linewidth]{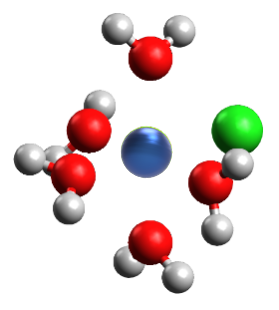}
    \newline
    Representative distorted M$^{q}$(H$_2$O)$_5$Cl$^{-}$ structure

    \caption{Structures of the model systems. 
    Top panel: The symmetric M$^{q}[\text{H}_2\text{O}]_6$ structure, consisting of a metal cation M$^{q}$ (blue) surrounded by six water molecules in a symmetric $D_{2h}$  structure, where the metal cation M$^{q}$ is located at the origin, and with two water molecules in each Cartesian direction.
    Middle panel: The disordered M$^{q}[\text{H}_2\text{O}]_6$ structure obtained from Molecular Dynamics, used as a representative structure for the dynamically disordered structure in the solution.
    Bottom panel:  The structure of M$^{q}[\text{H}_2\text{O}]_5\text{Cl}^-$, obtained by replacing one water molecule with a chloride ion (green), used as a representative structure for the case of ion pairing.
  }
    \label{fig:M_H2O_6_model_structure}
\end{figure}

Fig. \ref{fig:M_H2O_6_model_structure} shows schematic illustrations of the structures of the used model systems, symmetric M$^{q}[\text{H}_2\text{O}]_6$ structure (top), representative distorted M$^{q}$(H$_2$O)$_6$ structure (middle), and representative distorted M$^{q}$(H$_2$O)$_5$Cl$^{-}$ structure (bottom); the last two are average structures deriving from MD simulations for the aqueous solutions.

\subsection{Hybrid orbitals for the symmetric M$^{q}$[H$_2$O]$_6$ model system}
\begin{figure}[!htbp]
    \centering
    \includegraphics[width=0.8\textwidth]{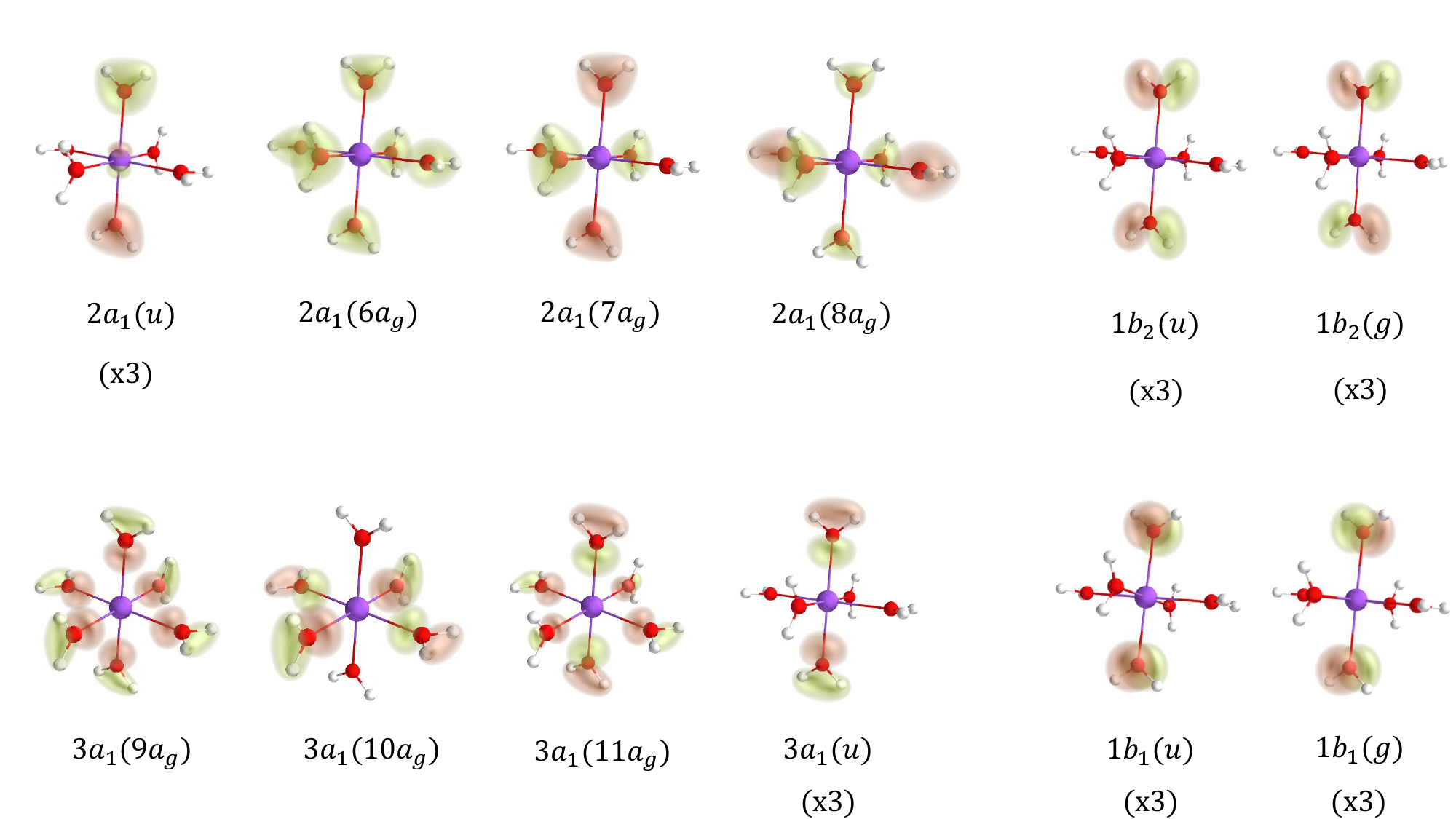}
    \caption{Hybrid orbitals for the symmetric M$^{q}[\text{H}_2\text{O}]_6$ model system.}
    \label{fig:na_panel_orb}
\end{figure}

Schematic illustrations of the calculated hybrid orbitals for the symmetric M$^{q}[\text{H}_2\text{O}]_6$ model system are shown in Fig. \ref{fig:na_panel_orb}.

\subsection{Mulliken charges}

In Table \ref{tab:Mulliken_table} the Mulliken charges for the different electronic states of Na and Mg are shown.
\begin{table}[!htpb]
    \setlength{\tabcolsep}{1pt}
    \renewcommand{\arraystretch}{1.5}
    \centering
    \caption{Mulliken charges for each electronic state per atomic center, for both $\text{M}[\text{H}_2\text{O}]_6$, M$=$Na, Mg systems. Each of the oxygen lines represents a symmetric pair of oxygen atoms, whilst each of the hydrogen lines represents four hydrogen atoms.}
    \begin{tiny}
    \begin{tabular}{|l|c|c|ccccc||c|c|ccccc|}
    \hline
    Atom&\multicolumn{7}{c||}{Na} & \multicolumn{7}{|c|}{Mg}\\
    \hline
    & ground &core hole&\multicolumn{5}{c||}{final state} & ground &core hole &\multicolumn{5}{|c|}{final state}\\
    \hline
    & & (M1$s)^{-1}$ & (M2$p)^{-1}$ & (2$a_1)^{-1}$ & (1$b_2)^{-1}$ & (3$a_1)^{-1}$ & (1$b_{1})^{-1}$ & & (M1$s)^{-1}$ &(M2$p)^{-1}$ & (2$a_1)^{-1}$ & (1$b_2)^{-1}$ & (3$a_1)^{-1}$ & (1$b_{1})^{-1}$ \\
    \hline
    M& 0.95 & 1.48 & 1.53 & 1.20 & 0.87 & 0.87 & 0.86 & 1.35 & 1.75 & 2.20 & 1.22 & 1.21 & 1.25 & 1.21 \\
    \hline
    O(1)& -0.42 & -0.46 & -0.48 & -0.48 & -0.16 & -0.48 & -0.48 & -0.49 & -0.47 & -0.53 & -0.53 & -0.20 & -0.53 & -0.53\\
    O(2)& -0.42 & -0.46 & -0.39 & -0.26 & -0.51 & -0.08 & -0.50 & -0.49 & -0.47 & -0.56 & -0.17 & -0.56 & -0.16 & -0.55 \\
    O(3)& -0.42 & -0.46 & -0.50 & -0.50 & -0.50 & -0.49 & -0.02 & -0.49 & -0.47 & -0.56 & -0.56 & -0.56 & -0.54 & -0.08 \\
    H(1)& 0.21 & 0.27 & 0.25 & 0.25 & 0.34 & 0.25 &  0.25 & 0.30 & 0.34 & 0.33 & 0.33 & 0.41 &  0.33 &  0.33 \\
    H(2)& 0.21 & 0.27 & 0.29 & 0.31 & 0.27 & 0.29 &  0.27 & 0.30 & 0.34 & 0.35 & 0.40 & 0.35 &  0.37 &  0.35 \\
    H(3)& 0.21 & 0.27 & 0.26 & 0.26 & 0.26 & 0.26 &  0.27 & 0.30 & 0.34 & 0.34 & 0.34 & 0.34 &  0.35 &  0.35 \\
    \hline
    \end{tabular}
    \end{tiny}
    \label{tab:Mulliken_table}
\end{table}

\subsection{The IRD mechanism}

\subsubsection{Symmetric structural model}

Here we  take the discussion of the IRD process further than in the main text by analyzing what mechanism provides the intensity of the IRD transitions using the symmetric structural model shown in the top panel of Fig. \ref{fig:M_H2O_6_model_structure}. 
The spectral intensity of transitions from the different hybrid orbitals ${\psi}_{\mu}$, for a given $\mu$, to the M \textit{1s} hole depends on the set of $C_\mu^\nu$ and the dipole transition moment operator resolved by the atomic orbitals basis set $\textbf{t}$.  
The intensity is then proportional to $|\textbf{T}_\mu|^2={\lvert\sum_\nu}$$\textbf{t}_{M1s,\nu}$${C}_{\mu}^\nu\rvert^{2}$, i.e., the squared sum of the products between the respective dipole transition moments $\textbf{t}_{\text{M}1s,\nu}$ = $\langle\chi_{\text{M}1s}\lvert \textbf{r}\rvert\chi_\nu\rangle$, in which \textbf{r} is the dipole operator, and the orbital coefficients, ${C}_\mu^\nu$. 

The IRD lines are due to decay from three different hybrid orbitals ${\psi}_{\mu}$, which consist mainly of ${1b}_{2}$, ${3a}_{1}$, and ${1b}_{1}$ water outer-valence molecular orbitals, respectively. Of the three IRD lines, the one from ${3a}_{1}$ has $>$10 times higher intensity than the ones from ${1b}_{2}$ and ${1b}_{1}$, see Table \ref{tab:Table_OCA_SI_short}, and we will consequently focus our discussion on this dominant ${3a}_{1}$ line.

In Table \ref{tab:Na_Mg_OCA_short_alternative2}, we present the main atomic orbital ${C}_\mu^\nu$ contributions to the hybrid orbitals ${\psi}_\mu$, the symmetry-allowed Cartesian component of $\textbf{t}_{\text{M}1s,\nu}$ - simply denoted by $\text{t}_\nu$ hereafter - as well as their product $\text{t}_\nu$${C}_\mu^\nu$. Note that in the present structural model only ungerade hybrid orbitals contribute to the spectra for symmetry reasons. Similarly, the M ${ns}$ contributions to the hybrid orbitals also do not contribute as $\textbf{t}_{\text{M}1s,\text{M}{ns}}=0$ due to dipole selection rules.

\begin{table}[!htpb]
    \setlength{\tabcolsep}{4pt}
    \renewcommand{\arraystretch}{1.5}
    \centering
    \caption{The main contributions to the spectral intensity for K$_{\alpha}$ and ${3a}_{1}$ IRD for the symmetric M$^q[\text{H}_2\text{O}]_6$ model, with M$^q$ = Na$^{+}$ and Mg$^{2+}$, in atomic units. The considered value of $R$ was $2.3$ \AA{} for Na and $2.1$ \AA{} for Mg. In the columns, t$_\nu$ denotes the respective symmetry-allowed Cartesian component of the atomic dipole transition moments $\textbf{t}_{\text{M}1s,\nu}$ and ${C}_\mu^\nu$ denotes the orbital coefficient of the atomic orbital $\nu$ in the LCAO description of the hybrid orbitals ${\psi}_{\mu}$. In the bottom row,  $|\textbf{T}_\mu|^2$ =  $|\sum_\nu\textbf{t}_\nu C_\mu^\nu|^2$ is proportional to the total spectral intensity of the K$_{\alpha}$ and ${3a}_{1}$ IRD transitions.}    
    \begin{footnotesize}
    \begin{tabular}{|l|c|cccc||c|cccc|}
    \hline
    Metal atom&\multicolumn{5}{c||}{Na} & \multicolumn{5}{|c|}{Mg}\\
    \hline
 Transition& & \multicolumn{2}{c}{K$_{\alpha}$}& \multicolumn{2}{c||}{IRD ${3a}_{1}$}& & \multicolumn{2}{c}{ K$_{\alpha}$}& \multicolumn{2}{c|}{IRD ${3a}_{1}$}\\
 & $\text{t}_\nu$ & $C_{\text{M}2p}^\nu$& $\text{t}_\nu C_{\text{M}2p}^\nu$& $C_{3a_1}^\nu$& $\text{t}_\nu C_{3a_1}^\nu$& $\text{t}_\nu$& $C_{\text{M}2p}^\nu$& t$_\nu C_{\text{M}2p}^\nu$& $C_{3a_1}^\nu$&t$_\nu C_{3a_1}^\nu$\\
    \hline
    Atomic orbital $\nu$ & {$\times10^{-3}$} & &$\times10^{-3}$& & $\times10^{-3}$& $\times10^{-3}$ & & $\times10^{-3}$& & $\times10^{-3}$\\
    \hline
    M 2p& 49.11& 0.97  &47.6&0.08 & 3.93& 47.05 & 0.99 & 46.6& 0.03 & 1.41\\
    M 3p&  -4.53& 0.01 & -0.045&-0.08 & 0.362& -6.78 & -0.01 & 0.068& -0.35& 2.37\\
    M 4p&  0.45& -0.02 & -0.009& 0.11 & 0.049& 0.93 & 0.01 & 0.009& 0.18& 0.167\\
    \hline
    O 2p& 0.04& 0.00& 0.002& 0.81 &  0.032& 0.06 & 0.00& 0.047& 0.79 & 0.047\\
    \hline
    \multicolumn{2}{|c|}{$|\textbf{T}_\mu|^2 \times10^{-6}$} &  \multicolumn{2}{c}{2405.99}& \multicolumn{2}{c||}{17.39}&  & \multicolumn{2}{c}{2216.23}& \multicolumn{2}{c|}{14.74}\\

    \hline
    \end{tabular}
    \end{footnotesize}
    \label{tab:Na_Mg_OCA_short_alternative2}
\end{table}

Table \ref{tab:Na_Mg_OCA_short_alternative2} forms a basis to discuss the mechanisms behind the different radiative transitions filling the M \textit{1s} core hole. The local K$_{\alpha}$ emission intensity is due to decay from a hybrid orbital with $\sim96\%$ M \textit{2p} character, making this an essentially atomic orbital centered on the metal ion. For both Na and Mg, the corresponding transition moment $\text{t}_{\text{M}2p}$ is $5-10$ times higher than any of the other transition moments, resulting in a very high intensity of the K$_{\alpha}$ line.

In contrast to the K$_{\alpha}$ case, the IRD  is due to decay from hybrid orbitals which are strongly dominated by O \textit{2p}, with much smaller M \textit{np} contributions. For the ${3a}_{1}$ IRD line, the corresponding $\psi_\mu$ has a  water character  of $\sim98\%$  for Na and $\sim85\%$ for Mg. The lower water character for Mg is consistent with the above discussed higher ground-state charge transfer to Mg than to Na. The charge transfer to Mg results in a partial population of orbitals derived from \textit{3s} and  \textit{3p}, which are spatially more extended than the  \textit{2p} and therefore hybridize stronger with water. The final ${3a}_{1}^{-1}$ state after the IRD thus primarily consists of a hole delocalized on the water molecules around the metal ion. 

As discussed above, the spectral intensity of decay from ${\psi}_\mu$ is given by $|\textbf{T}_\mu|^2$. For ${3a}_{1}$ of both Na and Mg, the by far largest term in the sum is due to M \textit{np} contributions, being roughly two orders of magnitude larger than the ones related to O \textit{nl}. 
The different M $np$ contributions differ in the sense that M \textit{2p} is fully occupied already in the free ions, while M \textit{3p} and M \textit{4p} are partially occupied in the solvated ion due to hybridization with the water molecular orbitals. For Na, the \textit{2p} contribution is 10 times higher than any other, whereas for Mg the term associated to Mg \textit{3p} is somewhat higher than for Mg \textit{2p}. This is consistent with the above discussed larger ground-state charge transfer from water to the metal ion for Mg$^{2+}$ than for Na$^{+}$, resulting from the increased ion-water hybridization with decreasing ion-water distance. For Na, the intensity thus comes from Na \textit{2p} - O \textit{2p} hybridization, while for Mg also ground state charge transfer into Mg \textit{3p} is important. The intensity in IRD is therefore mainly due to transitions from the ${\chi}_{M{np}}$ components of the hybrid orbitals. The small metal character of the ${\psi}_\mu$ is compensated by the atomic transition moments being much larger for the metal ion orbitals than for the water orbitals. 

We conclude that, for IRD, the intensity and the electron come from different sources. Even though the hybrid orbitals have relatively small metal character, the associated high transition moments result in a much higher contribution to the intensity than from the dominant oxygen character. At the same time, the final-state hole is delocalized on the nearest water molecules, meaning that the electron filling the M \textit{1s} core hole effectively mainly comes from the solvation shell water. In spite of  this non-local character of IRD, the strong contribution to the intensity from the small metal character of the hybrid orbitals makes the one-center approximation remain a good model, as further discussed below.

The above discussion is based on calculations using an idealized, highly symmetric structure, which is useful to capture the essentials of the IRD mechanism, but it may introduce a bias such as the prediction of very low intensity of the IRD bands deriving from the $1b_1$ and $1b_2$ orbitals of the water molecules. We will now proceed to a discussion of spectral contributions similar to the one presented above, but for the cluster models suggested by MD.

\subsubsection{Dynamically disordered structural model}
 The calculations presented here were performed for Na$^{+}$[H$_{2}$O]$_{6}$ and  Mg$^{2+}$[H$_{2}$O]$_{6}$ representative clusters derived from molecular dynamics calculations which indicate that the most typical configuration for the first solvation shell is made up of 6 water molecules for both cations, see the middle panel of Fig. \ref{fig:M_H2O_6_model_structure}. The main difference is the average distances between the cation and oxygen atoms of water,  2.4 Å for Na and about 2.0 Å for Mg, derived from our MD simulations. 

The HF method with the Ahlrichs-VTZ basis set was employed to describe both the initial M \textit{1s}$^{-1}$ state of the X-ray emission process, and the final states with frozen holes in the inner-valence and valence orbitals of M$^{q}$[H$_{2}$O]$_{6}$ clusters. 
The energies of the XES spectrum were obtained as the energy differences between the relaxed M \textit{1s}$^{-1}$ core binding energy and the frozen valence binding energies, while the spectral intensities were obtained from the dipole approximation for the spontaneous emission process. With the complete basis set of the M$^{q}$[H$_{2}$O]$_{6}$ cluster, the ground state of the single M$^{q}$ cation was also obtained, in order to evaluate the overlap between occupied orbitals of the M$^{q}$ cation and the occupied orbitals of the cluster.

\begin{figure}[!htbp]
    \centering
    \includegraphics[width=0.75\linewidth]{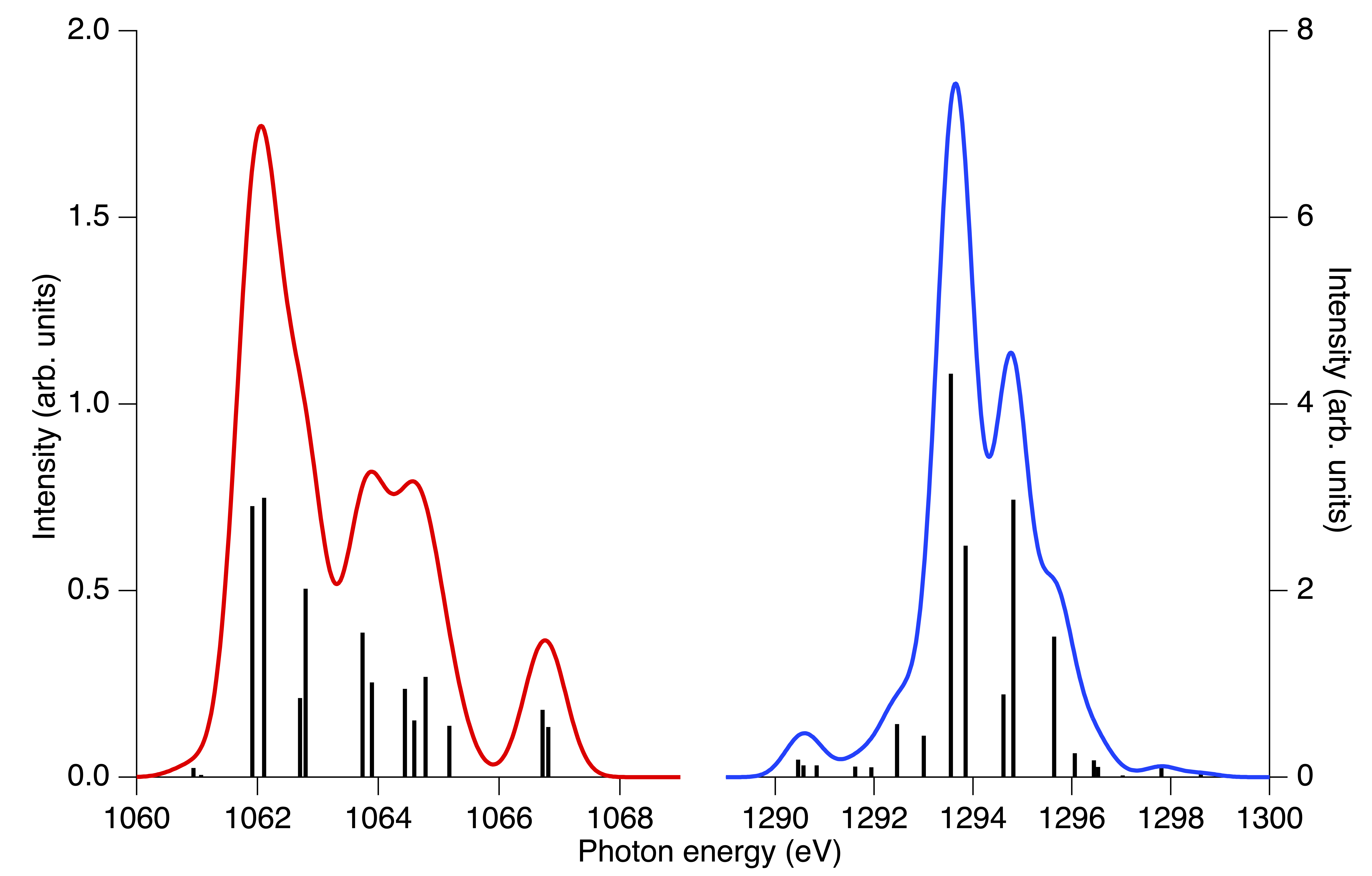}    
    \caption{XES-HF spectrum of the average clusters Na$^{+}$[H$_{2}$O]$_{6}$ (left) and Mg$^{2+}$[H$_{2}$O]$_{6}$ (right).}
    \label{fig:Na_Mg_IRD_VC}
\end{figure}

The computed XES spectra, limited to the high energy band, are shown in Fig. \ref{fig:Na_Mg_IRD_VC}. Original theoretical data are represented by black bars (intensity in arbitrary units), with the solid lines show their convolution with a Gaussian (FWHM=0.8 eV) to facilitate the comparison with the experimental spectra. 

In the case of decay of the M \textit{1s}$^{-1}$ core hole of the ion in water it is interesting to consider the electronic parameters that define the two characteristics of the spectrum, namely the energy position of the bands and their intensity. Since the core hole is extremely localized and of \textit{s} symmetry, the valence orbital involved in the decay process must have, in the dipolar approximation, a \textit{p}-type component on the cation site in order to provide an appreciable transition moment and thereby intensity.

It can be imagined that this can be achieved through two mechanisms:

\begin{enumerate}
    
\item "Hybridization" of the M \textit{2p} orbitals of the cation that are occupied in the isolated cation with occupied orbitals of the water molecules that surround it, so that these acquire a small  M \textit{2p} component on the cation site.

\item Population of the virtual M \textit{nl} orbitals (\textit{3s, 3p},…) of the isolated cation by electron transfer from the surrounding water molecules.
\end{enumerate}

Although mechanism 2 may seem particularly effective because it appears as an easy transfer of electrons towards a cation, it should be noted that this would mainly be to the M \textit{3s} orbital, whose population (despite the polarization due to the molecules of the solvent) cannot lead to an appreciable transition moment for the decay towards the core hole due to the dipole selection rule. The population of the virtual M \textit{3p}, to which the charge transfer would probably be smaller, could give a transition moment, but certainly lower than that of the M \textit{2p} orbital due to the presence of a nodal plane in the M \textit{3p} orbital. Mechanism 1 on the other hand involves a transition moment comparable to that of atomic decay.

The position of the XES bands evidently depends on the energies of the orbitals occupied in the condensed phase, i.e. the cluster in the present model. The orbitals occupied in the isolated cation remain occupied also in the cluster and their energy will shift slightly due to the polarization induced by the solvent molecules; the relative bands similar to those of the cation ("atomic" bands) remain strong and easily identifiable. The position of the additional XES bands depends on the remaining energy levels occupied in the cluster, i.e. those essentially localized on the water molecules. The latter, as the number of molecules increases, form pseudo-bands; that deriving from the 2a$_1$ orbital of the water molecule quite distinct from those deriving from the 3a$_1$, \textit{1b$_2$} and \textit{1b$_1$} orbitals, which practically form a single broad pseudo-band in which it is difficult to trace the components in terms of water orbital levels. Given the common geometrical structure of the first solvation shell for both Mg$^{2+}$ and Na$^{+}$, one can expect that the energetic bands (the occupied energy levels) due to the water molecules are comparable in the two cases. What can differ, is their relative position with respect to the "atomic" bands and their different contribution to the shape of the band depending on the specific transition moment.

To analyze this aspect it may be useful to consider Tables \ref{tab:Na_VC_long} and  \ref{tab:Mg_VC_long}, in which the following data are collected for the most intense XES bands calculated for both the Na$^{+}$[H$_{2}$O]$_{6}$ and  Mg$^{2+}$[H$_{2}$O]$_{6}$ clusters:

\begin{itemize}
    
\item in column 1 the index “i” of the occupied orbital of the cluster involved in the decay;

\item in column 2 the energy (eV) of the transition and in column 3 the corresponding intensity (arbitrary units); 

\item in columns 4, 5 and 6 the (percentage) overlap between the i orbital of the cluster and the three 2p orbitals occupied in the isolated cation; 

\item in columns 7, 8, 9 and 10 the overlap with the virtual orbitals \textit{3s} and \textit{3p} in the isolated cation; 

\item in the last column labels that identify the origin of the bands by grouping the single contributions.
\end{itemize}

\begin{table}[]
\begin{tiny}
\begin{tabular}{| c | cc | ccc | c | ccc | l|}
\hline
Na  & \multicolumn{2}{|c|}{$\text{t}_{\text{Na}{nl}}$} &   \multicolumn{3}{|c|}{Occupied in free ion}&        \multicolumn{4}{|c|}{Unoccupied in free ion}&                             \\
\hline
& & &         \multicolumn{3}{|c|}{4.9$\times10^{-2}$}&        0&                \multicolumn{3}{|c|}{-0.45$\times10^{-2}$}&            \\

i                              & E(ev)                                    & Int.                                                 & 2px     & 2py     & 2pz     & 3s     & 3px            & 3py      & 3pz    &                   \\
& & &         &         &         &        &                &          &        &                   \\
9                              & 1005.54                                  & 34.98                                                & 0.0029  & 0.0018  & 0.0000  & 0.0001 & 0.0000         & 0.0000   & 0.0000 & Na 2s \\

 & & & & & & & & & & \\

10                             & 1040.17                                  & 111.08                                               & 0.6251  & 31.9217 & 48.2016 & 0.1653 & 2.3422         & 0.0157   & 0.4290 & \multirow{3}{*}{\(\left\}\begin{array}{c} \\ \\ \\   \end{array}\right.\) Na 2p} \\
11                             & 1040.18                                  & 126.92                                               & 91.3305 & 0.3948  & 0.1176  & 0.2823 & 0.0045         & 0.7066   & 0.5833 & \\
12                             & 1040.23                                  & 127.92                                               & 0.0016  & 52.8804 & 39.7470 & 0.4699 & 0.0692         & 0.5102   & 0.3740 & \\

 & & & & & & & & & & \\

13                             & 1040.60                                  & 25.48                                                & 1.5698  & 13.3978 & 2.1968  & 5.1544 & 3.9905         & 1.9824   & 1.1546 & \multirow{5}{*}{\(\left\}\begin{array}{c} \\ \\ \\ \\ \\  \end{array}\right.\) O 2s} \\
14                             & 1040.71                                  & 19.28                                                & 5.1063  & 0.7134  & 7.0168  & 2.3414 & 2.4352         & 5.5708   & 3.2924 & \\
15                             & 1041.09                                  & 1.55                                                 & 0.4001  & 0.2310  & 0.3417  & 3.1137 & 2.0871         & 0.1462   & 6.6134 & \\
16                             & 1041.33                                  & 5.68                                                 & 0.8851  & 0.3962  & 2.1875  & 3.1351 & 5.4447         & 0.1375   & 7.1068 & \\
17                             & 1043.38                                  & 0.50                                                 & 0.0626  & 0.0244  & 0.1544  & 4.0484 & 1.9921 & 11.1222&        0.0991&                   \\

 & & & & & & & & & &\\

26                             & 1061.91                                  & 0.72                                                 & 0.0001  & 0.0072  & 0.0003  & 3.6641 & 0.0044         & 0.0425   & 0.1518 & \multirow{12}{*}{\(\left\}\begin{array}{c} \\ \\ \\ \\ \\ \\ \\ \\ \\ \\ \\ \\ \end{array}\right.\) O 2p} \\
27                             & 1062.11                                  & 0.74                                                 & 0.0005  & 0.0003  & 0.0109  & 1.0658 & 0.5840         & 0.0921   & 0.0206 & \\
28                             & 1062.70                                  & 0.21                                                 & 0.0007  & 0.0007  & 0.0013  & 0.6003 & 0.1864         & 0.6814   & 0.0275 & \\
29                             & 1062.79                                  & 0.50                                                 & 0.0047  & 0.0014  & 0.0030  & 0.3864 & 0.0716         & 0.1486   & 0.0069 & \\
30                             & 1063.73                                  & 0.38                                                 & 0.0014  & 0.0007  & 0.0004  & 1.1598 & 0.0836         & 0.4895   & 1.0501 & \\
31                             & 1063.89                                  & 0.25                                                 & 0.0006  & 0.0012  & 0.0011  & 0.4570 & 0.0057         & 0.0243   & 0.6185 & \\
32                             & 1064.44                                  & 0.23                                                 & 0.0002  & 0.0005  & 0.0016  & 0.2090 & 0.0002         & 0.2868   & 0.0037 & \\
33                             & 1064.59                                  & 0.15                                                 & 0.0017  & 0.0000  & 0.0005  & 0.5516 & 0.1596         & 0.4970   & 0.0015 & \\
34                             & 1064.78                                  & 0.26                                                 & 0.0010  & 0.0009  & 0.0015  & 1.2461 & 0.0558         & 0.7209   & 0.2650 & \\
35                             & 1065.17                                  & 0.13                                                 & 0.0002  & 0.0000  & 0.0009  & 0.0303 & 0.0884         & 0.0587   & 0.0968 & \\
36                             & 1066.71                                  & 0.18                                                 & 0.0005  & 0.0002  & 0.0010  & 0.7013 & 0.3459         & 0.1802   & 0.0277 & \\
37                             & 1066.81                                  & 0.13                                                 & 0.0000  & 0.0002  & 0.0002  & 0.3022 & 0.6346         & 0.1482   & 0.0141 &    \\
\hline
\end{tabular}
\end{tiny}
\caption{Contributions to the XES spectrum calculated for the average cluster Na$^{+}$[H$_{2}$O]$_{6}$ cluster. 
    Row 2: Atomic transition moments $\text{t}_{\text{Na}{nl}}$
    Column 1: The index “i” of the occupied orbital of the cluster involved in the decay. Column 2: The energy of the transition in eV. 
    Column 3: The corresponding  intensity (arbitrary units); 
Columns 4, 5, and 6: The (percentage) overlap between the i orbital of the cluster and the three \textit{2p} orbitals of the isolated cation. 
Columns 7, 8, 9, and 10: The overlap between the i orbital of the cluster and the virtual orbitals \textit{3s} and \textit{3p} of the isolated cation.}
 \label{tab:Na_VC_long}
\end{table}

\begin{table}[]
\begin{tiny}
\begin{tabular}{|c | cc | ccc | c | ccc | l|}
\hline
Mg & \multicolumn{2}{c}{$\text{t}_{\text{Mg}{nl}}$} & \multicolumn{3}{|c|}{Occupied in free ion}&         \multicolumn{4}{|c|}{Unoccupied in free ion}&                        Main source of electron\\

\hline

& & &          \multicolumn{3}{|c|}{4.7$\times10^{-2}$}&         0&        \multicolumn{3}{|c|}{-0.75$\times10^{-2}$}&                        \\
i                                                                                                                                              & E(eV)                                          & Int.                                                   & 2px      & 2py     & 2pz     & 3s      & 3px    & 3py    & 3pz     &                        \\
& & &          &         &         &         &        &        &         &                        \\
10                                                                                                                                             & 1211.03                                        & 4.09                                                   & 0.0002   & 0.0002  & 0.0001  & 0.0011  & 0.0000 & 0.0000 & 0.0000  & Mg 2s\\
 & & & & & & & & & &\\
11                                                                                                                                             & 1251.19                                        & 213.75                                                 & 9.8448   & 0.1161  & 90.0015 & 0.0000  & 0.0034 & 0.0023 & 0.0013  & \multirow{3}{*}{\(\left\}\begin{array}{c} \\ \\ \\   \end{array}\right.\) Mg 2p}  \\
12                                                                                                                                             & 1251.22                                        & 213.97                                                 & 26.9904  & 70.9727 & 2.0046  & 0.0003  & 0.0000 & 0.0017 & 0.0040  & \\
13                                                                                                                                             & 1251.25                                        & 214.11                                                 & 63.1319  & 28.8831 & 7.9577  & 0.0000  & 0.0019 & 0.0011 & 0.0010  &                        \\

 & & & & & & & & & &\\

14                                                                                                                                             & 1272.93                                        & 0.20                                                   & 0.0006   & 0.0007  & 0.0064  & 19.6584 & 4.4457 & 2.3805 & 0.4977  & \multirow{6}{*}{\(\left\}\begin{array}{c} \\ \\ \\ \\ \\ \\   \end{array}\right.\) O 2s}\\
15                                                                                                                                             & 1273.09                                        & 0.48                                                   & 0.0125   & 0.0035  & 0.0032  & 0.5195  & 2.0442 & 0.4152 & 21.4137 &                        \\
16                                                                                                                                             & 1273.29                                        & 0.26                                                   & 0.0002   & 0.0010  & 0.0091  & 1.9074  & 5.1259 & 8.1070 & 0.4751  &                        \\
17                                                                                                                                             & 1274.04                                        & 0.17                                                   & 0.0015   & 0.0023  & 0.0014  & 4.5710  & 1.8404 & 5.8810 & 6.8753  & \\
18                                                                                                                                             & 1274.41                                        & 0.16                                                   & 0.0017   & 0.0030  & 0.0005  & 2.0998  & 7.2954 & 5.6789 & 0.2516  &                        \\
19                                                                                                                                             & 1275.49                                        & 0.35                                                   & 0.0025   & 0.0057  & 0.0009  & 4.7862  & 7.5683 & 6.6547 & 0.1418  &                        \\

 & & & & & & & & & &\\
 
22                                                                                                                                             & 1290.46                                        & 0.18                                                   & 0.0000   & 0.0004  & 0.0000  & 0.0125  & 0.7614 & 1.5813 & 0.9184  & \multirow{15}{*}{\(\left\}\begin{array}{c} \\ \\ \\ \\ \\ \\ \\ \\ \\ \\ \\ \\ \\ \\ \\   \end{array}\right.\) O 2p} \\
23                                                                                                                                             & 1290.57                                        & 0.12                                                   & 0.0000   & 0.0000  & 0.0003  & 0.0654  & 1.3636 & 0.0673 & 0.0443 &                        \\
24                                                                                                                                             & 1290.83                                        & 0.12                                                   & 0.0000   & 0.0000  & 0.0003  & 0.2281  & 1.7801 & 0.6706 & 0.0000  &                        \\
25                                                                                                                                             & 1291.61                                        & 0.11                                                   & 0.0001   & 0.0000  & 0.0002  & 0.6950  & 0.3078 & 0.1568 & 1.7463  &                        \\
26                                                                                                                                             & 1291.94                                        & 0.10                                                   & 0.0002   & 0.0000  & 0.0000  & 0.3661  & 0.1404 & 1.1950 & 0.3565  &                        \\
27                                                                                                                                             & 1292.46                                        & 0.56                                                   & 0.0005   & 0.0004  & 0.0010  & 15.3684 & 0.2321 & 0.5804 & 0.0589  &                        \\
28                                                                                                                                             & 1293.00                                        & 0.44                                                   & 0.0000   & 0.0002  & 0.0008  & 0.7411  & 0.7571 & 1.2746 & 0.2765  &                        \\
29                                                                                                                                             & 1293.55                                        & 4.32                                                   & 0.0074   & 0.0011  & 0.0039  & 0.0537  & 0.0366 & 0.0070 & 1.4721  &                        \\
30                                                                                                                                             & 1293.84                                        & 2.48                                                   & 0.0007   & 0.0020  & 0.0043  & 0.2403  & 0.0236 & 0.2547 & 0.0638  &                        \\
31                                                                                                                                             & 1294.61                                        & 0.88                                                   & 0.0003   & 0.0004  & 0.0009  & 0.8138  & 0.0147 & 0.0456 & 0.3782  &                        \\
32                                                                                                                                             & 1294.81                                        & 2.97                                                   & 0.0014   & 0.0031  & 0.0009  & 0.0166  & 1.0001 & 0.5138 & 0.3829  &                        \\
33                                                                                                                                             & 1295.64                                        & 1.50                                                   & 0.0009   & 0.0015  & 0.0002  & 0.9406  & 0.2858 & 0.4912 & 0.0166  &                        \\
34                                                                                                                                             & 1296.06                                        & 0.25                                                   & 0.0001   & 0.0006  & 0.0000  & 0.0000  & 0.1656 & 0.7476 & 0.0059  &                        \\
35                                                                                                                                             & 1296.44                                        & 0.18                                                   & 0.0002   & 0.0002  & 0.0000  & 0.0158  & 0.5697 & 0.1356 & 0.3449  &                        \\
36                                                                                                                                             & 1296.53                                        & 0.10                                                   & 0.0000   & 0.0000  & 0.0003  & 0.0000  & 0.0812 & 0.0283 & 0.9153  & \\
\hline
\end{tabular}
\end{tiny}
\caption{Contributions to the XES spectrum calculated for the average cluster Mg$^{2+}$[H$_{2}$O]$_{6}$ cluster. 
    Row 2: Atomic transition moments $\text{t}_{\text{Mg}{nl}}$
    Column 1: The index “i” of the occupied orbital of the cluster involved in the decay. Column 2: The energy of the transition in eV. 
    Column 3: The corresponding  intensity (arbitrary units); 
Columns 4, 5, and 6: The (percentage) overlap between the i orbital of the cluster and the three \textit{2p} orbitals of the isolated cation. 
Columns 7, 8, 9, and 10: The overlap between the i orbital of the cluster and the virtual orbitals \textit{3s} and \textit{3p} of the isolated cation.}
 \label{tab:Mg_VC_long}
\end{table}

In the case of the  Mg$^{2+}$[H$_{2}$O]$_{6}$ cluster, see Table \ref{tab:Mg_VC_long}, the "atomic" bands are well identified and isolated (the Mg(2s) band, which would be dark in the atom, becomes visible due to the deformation of the spherical symmetry in the presence of water molecules). It is followed (+22 eV with respect to the "atomic" Mg(2p) band) by a weak, relatively narrow and well-resolved band, due to the 2a$_1$ orbitals of the water molecules. Finally, in the range (+42 eV -- +46 eV) there is a structured larger band due, generically, to the 3a$_1$/1b$_2$/1b$_1$ valence orbitals. The percentage overlaps with the occupied 2p orbitals of Mg$^{2+}$ clearly show how the most appreciable intensities for the secondary bands of the cluster correspond to appreciable overlap values ("hybridization" mechanism). While to the orbitals (for example 14, 25 and 27) which are not affected by this mechanism, despite showing the occurrence of a substantial transfer of charge towards the virtual 3s and 3p orbitals, correspond to more modest intensities.

In the case of the Na$^{+}$[H$_{2}$O]$_{6}$  cluster, see Table \ref{tab:Na_VC_long}, the effectiveness of the hybridization mechanism is confirmed by the percentage values of the overlap with \textit{2p} orbitals of Na$^+$ in the second table. At the same time the charge transfer towards the virtual \textit{3s} and \textit{3p} orbitals appears less relevant than in Mg$^{2+}$[H$_{2}$O]$_{6}$, in reasonable agreement with an increased distance between cation and water molecules. The main differences of the Na spectrum compared to that of Mg, can be related to the difference of the M \textit{2p} energy level in the two cations with respect to the energy position of the water pseudo-bands which are similar in the two cases. The HF calculation provides for the \textit{2p} level of Na$^+$ a value of about 41 eV which is particularly close (37 eV) to that of the \textit{2a$_1$} water orbital. This favors a strong mixing of such orbitals (see for example the overlap values for the cluster orbitals 13 and 14) with a consequently strong intensity of transitions which are energetically very close to the "atomic" band Na(\textit{2p}). The remarkable hybridization of the inner valence shell of the waters goes to the detriment of that with the valence shell bands, which in fact show overlap and intensity of lower value and with a decidedly wider energy distribution than in the case of Mg.

\subsection{IRD and the One Center Approximation}

X-rays are emitted as valence electrons fill atom-specific local core holes. This makes X-ray emission spectroscopy (XES) sensitive to the local electronic structure, and a merit of the method is that simplifying interpretation schemes can often be applied. Such schemes are typically based on the one-center model and the dipole selection rule \cite{Manne1970}. Here the XES intensity depends on the local angular-momentum character of the core hole and the valence states, e.g., a quasi-atomic 1s hole is filled by electrons from valence states of local p character only. The validity of this model is often taken for granted, and it considered that XES measures the local partial density of states.

Simple molecular models for Na$^{+}[\text{H}_2\text{O}]_n$ and Mg$^{2+}[\text{H}_2\text{O}]_n$ were investigated, with the purpose of better understanding the competitive mechanisms within the IRD processes. The choice of the number of molecules used in these cluster models was $n=6$ for both systems, consistent with the first solvation shells in aqueous solutions. For simplicity, a symmetric $D_{2h}$ structure has been considered, where the metal cation is located at the origin and with two water molecules disposed along each cartesian direction, see the upper panel of Fig. \ref{fig:M_H2O_6_model_structure}.  The resulting valence MO's are shown in Fig. \ref{fig:na_panel_orb} according to the different components of the point group of symmetry . The symmetry-adapted ANO-RCC-PVTZ basis set has been used, and we have labeled each equivalent pair of oxygen atoms as O(1), O(2), and O(3). The distances $R$ between the metal ion and the oxygen atoms are the same in all three directions and $R=2.3$ \AA{} and $R=2.1$ \AA{} have been used for Na$^{+}$ and Mg$^{2+}$, respectively, also motivated by the behavior of the first solvation shells.

In order to have a better view of the different mechanisms, we consider a simple molecular orbital approach. Taking a closed-shell HF reference for the $=\text{M}^{q}$$[\text{H}_2\text{O}]_6$ small models, where $\text{M}^{q}$$=\text{Na}^{+}$, $\text{Mg}^{2+}$ denotes the metal cation, the first-order X-ray emission amplitude from a particular MO (indexed by $\mu$) to the ion 1$s$ core hole is given by the dipole moment matrix element given by
\begin{equation}
    \textbf{T}_\mu=\langle\psi_{\text{M}1s}|\textbf{r}|\psi_\mu\rangle=\langle\psi_{\text{M}1s}|\textbf{r}|\psi^\text{M}_\mu + \psi^\text{W}_\mu\rangle=\textbf{T}^\text{OC}_\mu+\textbf{T}^\text{MC}_\mu,
\end{equation}
where the sum $\psi^\text{M}_\mu + \psi^\text{W}_\mu$ explicitly decomposes the MO according to contributions arising from the metal (M)- and the surrounding water molecules (W)-centered basis functions. This allows us to conceptually decompose $\textbf{T}_\mu$ into  a one-center (OC) part , $\textbf{T}^\text{OC}_\mu$, due to transitions from M-centered wavefunctions, and a multi-center (MC) part, $\textbf{T}^\text{MC}_\mu$, due to transitions from W-centered wave functions. Since the intensity is proportional to $|\textbf{T}_\mu|^2$, one can write
\begin{equation}
    f^{\text{IRD}}_\mu=f^{\text{OC}}_{\mu}+f^{\text{MC}}_{\mu}+f^{\text{I}}_{\mu}\propto |\textbf{T}^\text{OC}_\mu|^2 + |\textbf{T}^\text{MC}_\mu|^2+\left(\textbf{T}^\text{OC}_\mu\cdot\textbf{T}^\text{MC}_\mu+c.c.\right)=|\textbf{T}_\mu|^2,
\end{equation}
where $f^{\text{I}}_i$ is an interference term. 
Two approximations within the MO picture have been considered for the emission spectra. The first one is based on the evaluation of $|\textbf{T}_\mu|^2$ itself, called MO$\rightarrow$MO calculation. The second is the so-called one-center approximation (OCA), in which we approximate $|\textbf{T}_\mu|^2\sim|\textbf{T}^{\text{OC}}_\mu|^2$, i.e. neglecting all MC contributions. A comparison between these two approaches provides information on the magnitude of each mechanism to the IRD.

\begin{table}[!htpb]
    \setlength{\tabcolsep}{4pt}
    \renewcommand{\arraystretch}{1.5}
    \centering
    \caption{Cartesian components of the transition dipole moments ($\langle\text{M}1s|r_i|\chi_\nu\rangle$) resolved in the (symmetry adapted) AO basis set for M$[\text{H}_2\text{O}]_6$, for M=Na$^{1+}$, Mg$^{2+}$, in atomic units. The calculations were addressed using the aug-cc-PVTZ basis set. The considered value of $R$ was $2.3$ \AA{} for Na and $2.1$ \AA{} for Mg. The columns 3 to 12 display the MO coefficients for the 2$a_1$, 1$b_2$, 3$a_1$, and 1$b_1$ valence orbitals of the water molecules corresponding to each AO $\chi_\nu$. The emission energies in the last row were obtained from the RASSCF calculation (see text).}
    \label{tab:Table_OCA_SI_short}
    \begin{footnotesize}
    \begin{tabular}{|l|c|cccc||c|cccc|}
    \hline
    Metal atom&\multicolumn{5}{c||}{Na} & \multicolumn{5}{|c|}{Mg}\\
    \hline
    & $\times10^{-3}$ & 2$a_1(u)$ &1$b_2(u)$ & 3$a_1(u)$ & 1$b_{1}(u)$ & $\times10^{-3}$ & 2$a_1(u)$ & 1$b_2(u)$ & 3$a_1(u)$ & 1$b_{1}(u)$ \\
    \hline
    $\langle\text{M}1s|r_i|\text{M}2p_i\rangle$& 49.11& -0.27  &0.01&0.08 & 0.01 & 47.05 & -0.02 & -0.01 & 0.03 & 0.02 \\
    $\langle\text{M}1s|r_i|\text{M}3p_i\rangle$&  -4.53& -0.12 & -0.03&-0.08 & 0.18 & -6.78 & 0.10 & -0.18 & -0.35& 0.12 \\
    $\langle\text{M}1s|r_i|\text{M}4p_i\rangle$&  0.45& 0.05 & 0.01 & 0.11 & -0.10 & 0.93 & -0.12 & 0.18 & 0.18& -0.15 \\
    \hline
    $\langle\text{M}1s|r_i|\text{O}(1)2s\rangle$& 0& 0.84 & 0 & -0.28 &  -0.05 & 0 & 0.86& 0 & -0.23 & -0.07 \\
    $\langle\text{M}1s|r_i|\text{O}(1)2p_i\rangle$& 0.04& 0 & -0.74& 0.04 &  -0.01 & 0.06 & 0& -0.74 & 0.11 & -0.01 \\
    $\langle\text{M}1s|r_i|\text{O}(2)2p_i\rangle$& 0.04& 0.11&  0.04&0.81 & 0.09  & 0.06 & 0.10 & 0.09 & 0.79 & 0.14 \\
    $\langle\text{M}1s|r_i|\text{O}(3)2p_i\rangle$& 0.04& 0 & -0.01&-0.08 & 0.91  & 0.06 & 0 & -0.02 & -0.12 & 0.91 \\
    $\langle\text{M}1s|r_i|\text{O}(1)3p_i\rangle$& 0.00&  0 & 0.02&0.01 & 0 & 0.01 & 0 & 0.02 & 0 & 0 \\
    $\langle\text{M}1s|r_i|\text{O}(2)3p_i\rangle$& 0.00& 0.01& 0&-0.01 & 0 & 0.01 & 0 & 0 & -0.01 & 0 \\
    $\langle\text{M}1s|r_i|\text{O}(3)3p_i\rangle$& 0.00&  0& 0&0 & 0 & 0.01 & 0 & 0 & 0 & 0 \\
    \hline
    \hline
    $|\textbf{T}^\text{OC}_\mu|^2$ & $\times10^{-6}$  & 152.00 & 0.51 & 19.73 & 0.11 & $\times10^{-6}$ & 2.89 & 0.98 & 16.20 & 0.03 \\
    \hline
    $|\textbf{T}^\text{MC}_\mu|^2$& $\times10^{-6}$ & 0.01 & 0 & 0.07 & 0 & $\times10^{-6}$ & 0.02 & 0.01 & 0.03 & 0\\
    \hline
    $|\textbf{T}_\mu|^2$ & $\times10^{-6}$ & 150.06& 0.48 & 17.39 & 0.12 &$\times10^{-6}$ &2.41 & 0.80 & 14.74 & 0.01\\
    \hline
    \hline

    \multicolumn{2}{|l|}{Emission energy (eV)} & 1044.9& 1058.8& 1062.9 & 1065.2 & \multicolumn{2}{r}{1278.7} & 1293.6 & 1295.8 & 1299.9\\
    
    \hline
    \end{tabular}
    \end{footnotesize}
\end{table}

Table \ref{tab:Table_OCA_SI_short} summarizes the MO$\rightarrow$MO and the OCA calculations for the small model, by MO, showing individual contributions from atomic orbitals (AO) of the basis set and their respective MO coefficients. The $gerade$ MO's have zero transition moments, thus we only discuss the $ungerade (u)$ orbitals. Each symmetry-adapted 2$a_1(u)$, 1$b_2(u)$, 3$a_1(u)$ and 1$b_1(u)$ term denotes a set of three orbitals at once, associated with the $b_{1u}$, $b_{2u}$ and $b_{3u}$ components, in which the MO coefficients are the same, although interchangeable between O(1), O(2) and O(3) centers. AO's that have zero $\langle\text{M}1s|r_i|\chi_\nu\rangle$ elements are not shown, as well as the contributions from $2p$ orbitals on the hydrogen atoms. Furthermore, the OCA calculations contemplate $np\rightarrow1s$ atomic contributions for all $np$, $n=2,3,4,5$ on the metal ion, but only the most significant $n=2$ and $n=3$ contributions are shown in Table~\ref{tab:Table_OCA_SI_short} for simplicity.

\begin{figure}[!htbp]
    \centering
    \includegraphics[width=0.95\linewidth]{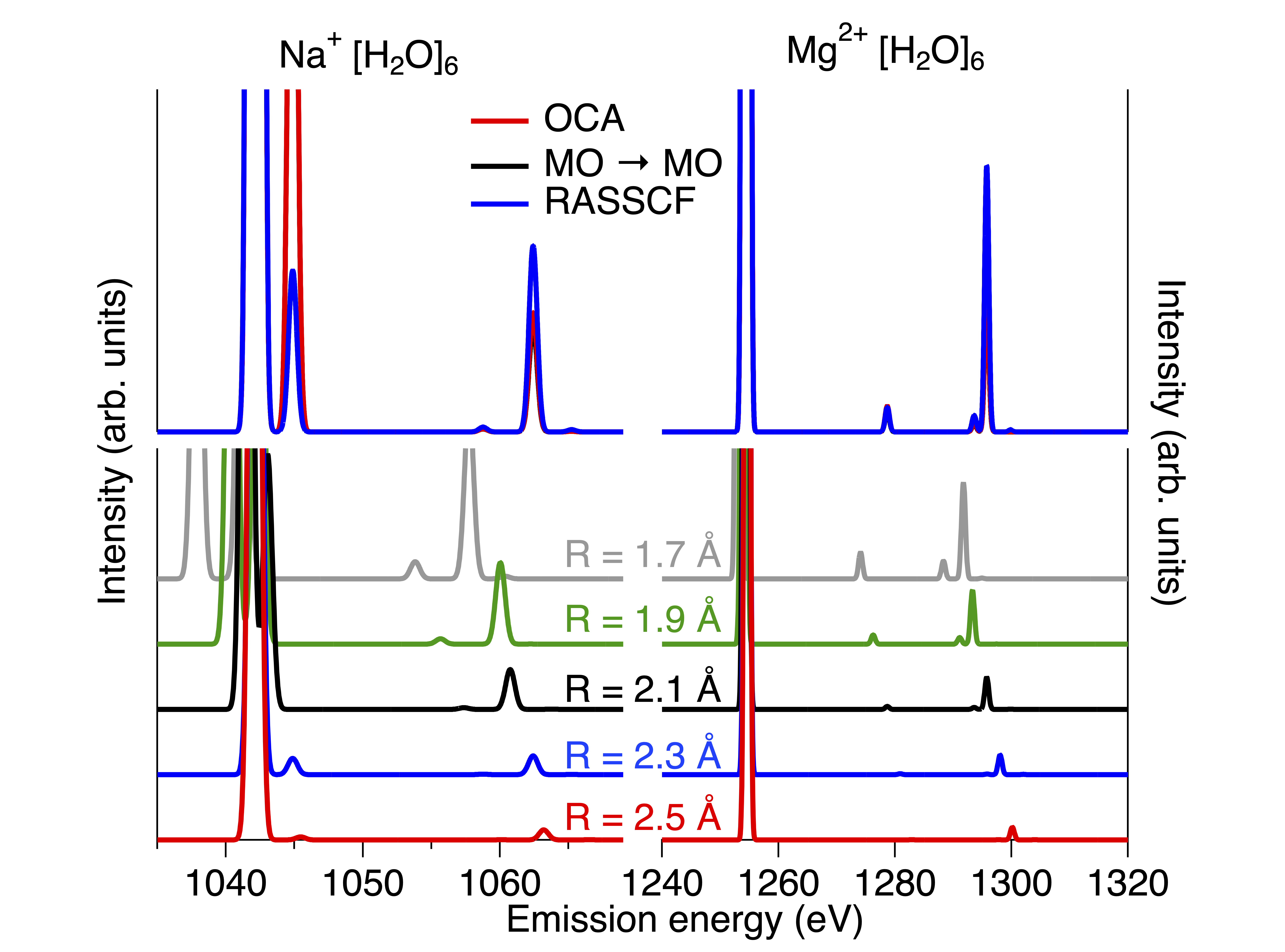}
    \caption{Top panel: Comparison between calculated XES spectra after M\textit{ 1s} ionization for Na$^{+}$[H$_{2}$O]$_{6}$ and  Mg$^{2+}$[H$_{2}$O]$_{6}$ obtained using OCA (red), MO $\rightarrow$ MO (black) and RASSCF (blue). Note that red and black curves overlap to a large extent.
    Lower panel: Calculated XES spectra after M 1s ionization for Na$^{+}$[H$_{2}$O]$_{6}$ and  Mg$^{2+}$[H$_{2}$O]$_{6}$ as function of M-water distance, R.}
    \label{fig:OCA_vs_MOMO}
\end{figure}

Comparing the values for $|\textbf{T}^\text{OC}_\mu|^2$, representing  OCA taking only contributions from metal ion orbitals, $|\textbf{T}^\text{MC}_\mu|^2$, representing cross-transitions from oxygen orbitals, and $|\textbf{T}_\mu|^2$, representing MO $\rightarrow$ MO, which includes contributions from both metal ion and oxygen orbitals, we see that the $|\textbf{T}^\text{MC}_\mu|^2$  values are much smaller than $|\textbf{T}^\text{OC}_\mu|^2$. 
This is illustrated in Fig. \ref{fig:OCA_vs_MOMO}, which shows that the spectra simulated using OCA and MO $\rightarrow$ MO are very similar in shape and intensity. The result from a RASSCF calculaion is also included, which agrees rather well with OCA and MO $\rightarrow$ MO regarding the spectral shape, but differs in IRD/K$_{\alpha}$ intensity ratio. 
This implies that the cross-transitions only weakly affect the total intensity, and that OCA describes the intensity quite well.

\subsection*{IRD Distance Dependence}
The efficiency of other non-local decay processes is known to decrease rapidly with increasing distance, for e.g., ICD, the efficiency is considered to scale with 1/R$^{6}$ for longer distances \cite{Jahnke2020}. The ability of IRD to probe the first solvation shell depends on the IRD efficiency decreasing rapidly with distance. 
We start by investigating the efficiency of IRD as a function of distance between the metal ion and the solvation shell water molecules by examining spectra computed using Hartree Fock Restricted Active Space Self-Consistent Field (RASSCF) methodology  (HF/RAS), shown in Fig. \ref{fig:XES(n,r)}, for Na$^{+}$ and Mg$^{2+}$ surrounded by one, four, and six water molecules in symmetric positions.  

\begin{figure}[!htbp]
    \centering
    \includegraphics[width=0.95\linewidth]{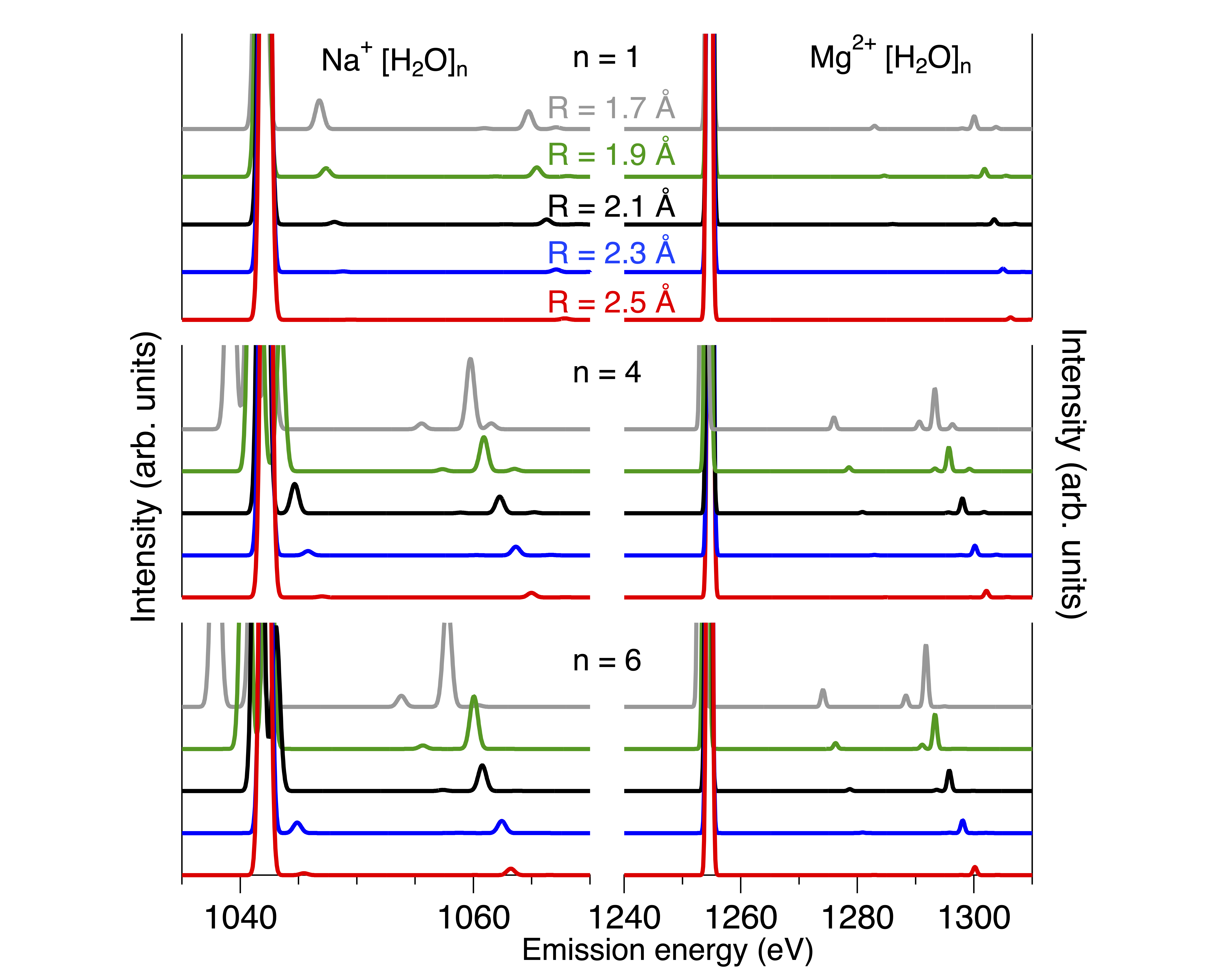}
  \caption{Calculated XES spectra after M 1s ionization for Na$^{+}$[H$_{2}$O]$_{n}$ and  Mg$^{2+}$[H$_{2}$O]$_{n}$ as function of M-water distance (R), and number of water molecules (n).}
    \label{fig:XES(n,r)}
\end{figure}

\begin{figure}[!htbp]
    \centering
    \includegraphics[width=0.8\linewidth]{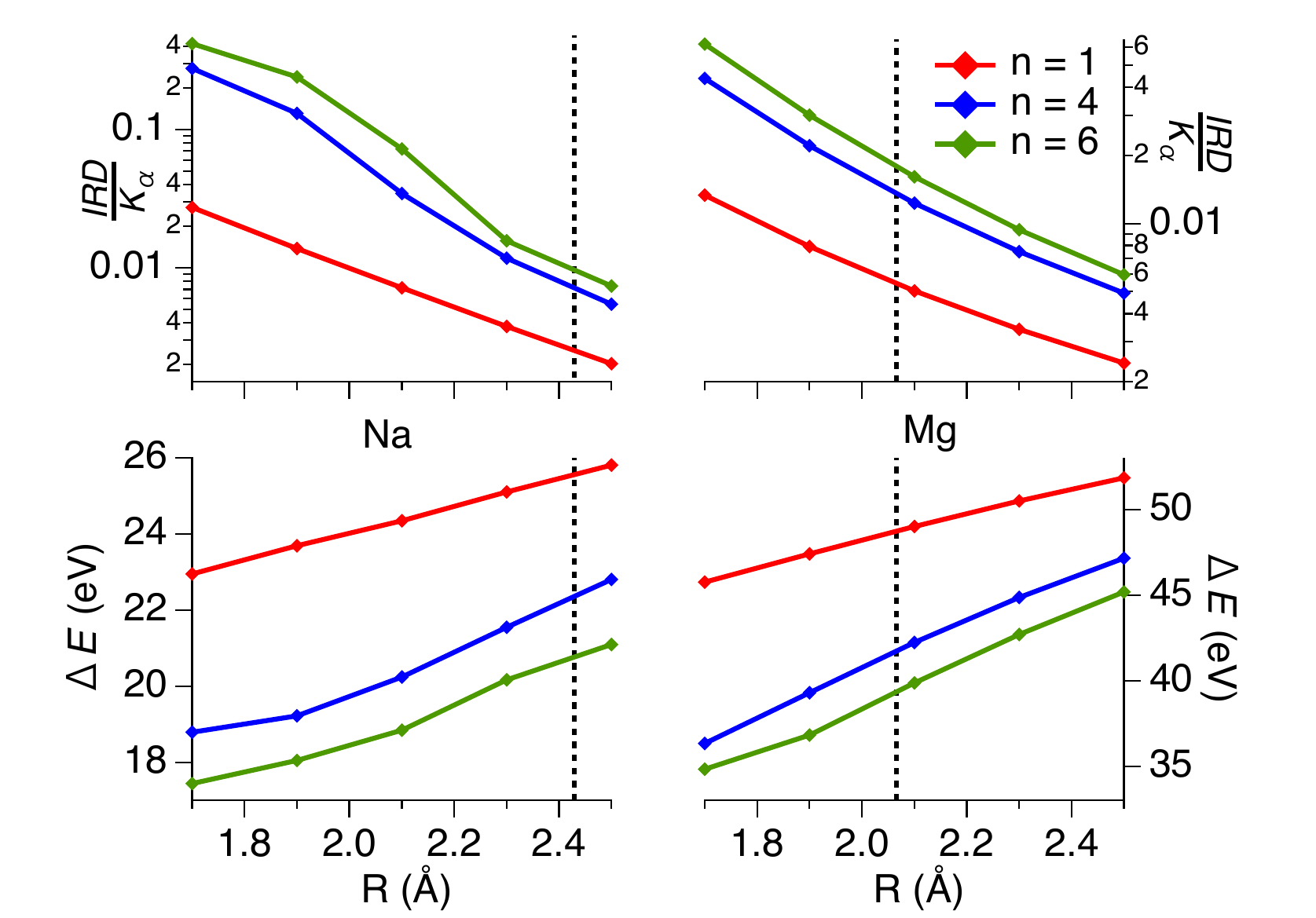}
\caption{Calculated IRD/K$_{\alpha}$ intensity ratio and IRD-K$_{\alpha}$ energy splitting $\Delta$E as a function of M-water distance R and number of water molecules \textit{n} for Na$^{+}$[H$_{2}$O]$_{n}$ (left) and  Mg$^{2+}$[H$_{2}$O]$_{n}$ (right). The vertical dashed lines correspond to  the experimentally determined M-O distances in solution, 2.43 Å and 2.07 Å for Na and Mg, respectively \cite{Persson2024}.}
    \label{fig:IRD_kalpha_ratio}
\end{figure}

From the calculations , the IRD/K$_{\alpha}$ intensity ratio  was derived, see Fig.  \ref{fig:IRD_kalpha_ratio}. As can be seen, the IRD/K$_{\alpha}$ intensity ratio increases with increasing number of water molecules, consistent with the IRD being caused by the solvation-shell water, and with decreasing distance. 

A fit of the distance dependence of the IRD/K$_{\alpha}$ intensity ratio for \textit{n}=1 to a 1/R$^{x}$ function yields x$\approx$6.4 for Na and x$\approx$4.5 for Mg. The difference between the two ions can be attributed to the different amounts of hybridization that occur as a function of distance and orbital size. As the number of water molecules in a shell of radius R around the metal ion scales with $\sim$R$^{2}$, the values obtained for the 1/R$^{x}$ distance dependence of the IRD/K$_{\alpha}$ intensity ratio imply that the first solvation shell will dominate the IRD process.

The dependence of the IRD/K$_{\alpha}$ intensity ratio on the number of neighboring water molecules opens the possibility of quantitatively analyzing the composition of the solvation shell, for example, in connection to ion pairing as discussed further below. For the experimentally determined M-O distances in solution, the calculated  IRD/\textit{K$_{\alpha}$ }ratio for \textit{n}=6 is $\sim$1.0 $\%$ and  $\sim$1.8 $\%$ for Na and Mg, respectively, to be compared to our experimental values of $\sim$1.0 $\%$ and  $\sim$1.5 $\%$. For Na the agreement is excellent, and for Mg an even better agreement was obtained with a more realistic structure model derived from MD simulations of the aqueous solution, asdiscussed in the main text. 

Parallel to the increase of the IRD/K$_{\alpha}$ intensity ratio with decreasing distance, the lower panels of  Fig. \ref{fig:IRD_kalpha_ratio}  shows how the energy splitting between the K$_{\alpha}$ and IRD spectral features, the latter defined  as the energy of the main  ${3a}_{1}$ peak, decreases with decreasing distance. This change can qualitatively be understood as being due to an increasing hybridization between orbitals on the ion and the water molecules with decreasing distance \textit{R}. For the experimentally determined M-O distances in solution, the energy splitting obtained from calculation for\textit{ n}=6 is $\sim$21 eV for Na and $\sim$39 eV for Mg, respectively, in fair agreement with our experimental values of $\sim$23 eV for Na and $\sim$41 eV for Mg.

We conclude that both the IRD/K$_{\alpha}$ intensity ratio and  energy splitting  are strongly sensitive to distance and number of nearest neighbors, pointing towards using IRD as a probe to explore the solvation shell.

\subsection{IRD and solvation shell (dis)order}

\newpage

In addition to energy shifts of the one-hole states, the spectra in Fig. \ref{fig:valence bands}  also exhibit differences in relative intensity between the  \textit{1b$_{1}$}, \textit{3a$_{1}$}, and \textit{ 1b$_{2}$} peaks. For both Na and Mg, \textit{1b$_{2}$} is much weaker than the other two. The \textit{1b$_{1}$} and \textit{3a$_{1}$} features, however, exhibit differences between Na and Mg. For Na, they have roughly equal intensity, whereas for Mg, \textit{3a$_{1}$} has higher intensity than\textit{ 1b$_{1}$}. As we will show, these differences can be understood as caused by the different local geometries of the first solvation shell around Na and Mg. 

The idealized orientation of the water molecules in the solvation shell is with the negatively charged oxygen pointing towards the M cation and the M-O direction located in the molecular plane, i.e., with the hydrogen atoms pointing away from the cation. In solution, dynamic disorder will cause the molecular orientation to differ from the idealized one. 

\begin{figure}[!htbp]
    \centering
    \includegraphics[width=0.75\linewidth]{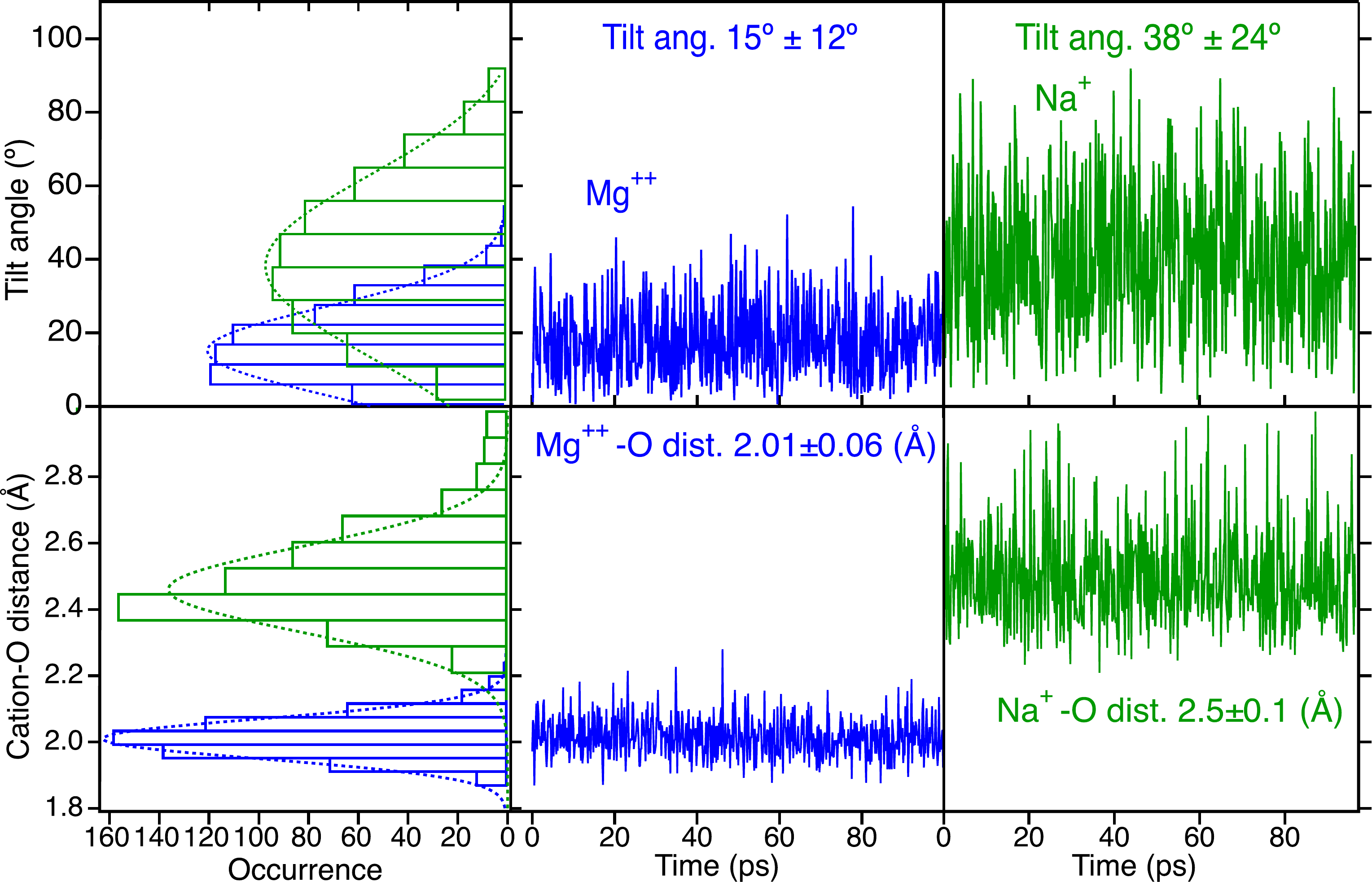}
    \caption{ The dynamic disorder of the  solvation shell water molecules around Na$^{+}$(green) and  Mg$^{2+}$ (blue ) obtained from the MD simulations.
    The lower row shows the M-O distance, the upper row the tilt angle. Examples of time traces from the MD simulation are shown in the middle and right columns. From the time traces,  distributions of distances and tilt angles were obtained, see the left column. The distribution of  distance and tilt angle values are shown as  histograms, to which Gaussian fits were made.  The given values for the distance and tilt angle are time averages ${\pm}$ the one sigma value from the Gaussian fits.   }
    \label{fig:Orientation from MD}
\end{figure}

 Based on our MD simulations, Fig. \ref{fig:Orientation from MD} illustrates the dynamic disorder in the solvation shell around Na$^{+}$ and Mg$^{2+}$ ions by showing time-traces of distance and orientation. 
 For Na$^{+}$ and Mg$^{2+}$, the time-averaged distance from MD is 2.5±0.1 Å, and 2.01±0.06 Å, respectively, in quite good agreement with the experimentally determined distances (2.43 Å and 2.07 Å, respectively \cite{Persson2024}).

\begin{figure}[!htbp]
    \centering
    \includegraphics[width=0.8\linewidth]{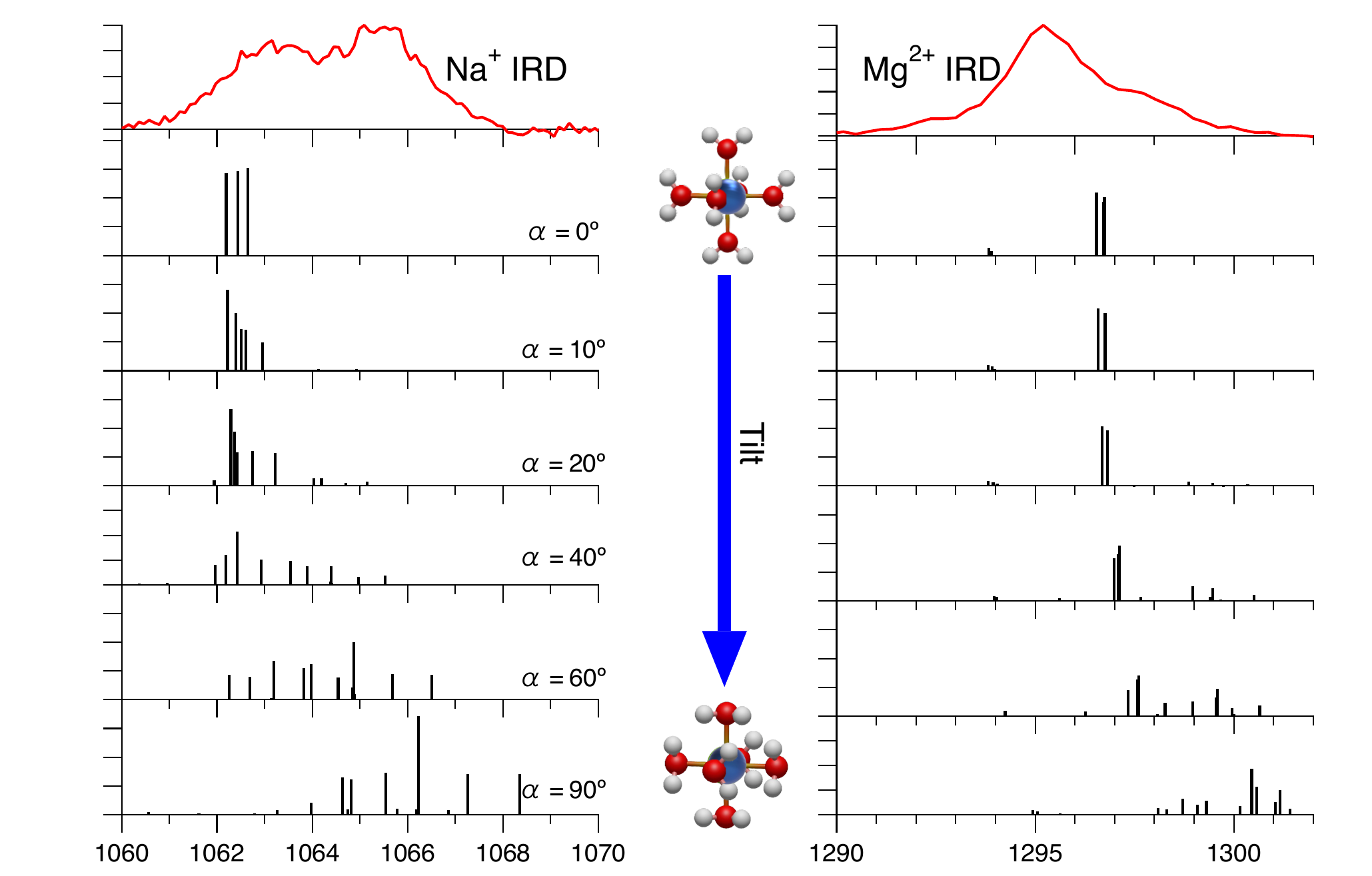}
    \caption{Calculated IRD spectra for Na$^{+}$[H$_{2}$O]$_{6}$ and Mg$^{2+}$[H$_{2}$O]$_{6}$ with various degrees of non-ideal structure (black vertical bars), compared to the experimental spectra (red curves). Tilt refers to the angle between the ion-oxygen radial vector and the dipole vector of the individual water molecules, with 0$^{\circ}$ being the idealized structure. The same energetic shift as in Fig. \ref{img:spectra} has been applied to the theoretical spectra.}
    \label{fig:orientation}
\end{figure}

We consider two simplified types of disorder, tilting and rotation. The tilt  is the angle $  \alpha$ between the dipole vector of the individual water molecules and the radial ion-oxygen vector, see  Fig. \ref{fig:orientation}. In all cases, the M-O distance was kept fixed at the average value provided by MD simulations, 2.3 and 2.0 Å for Na and Mg, respectively. The idealized orientation corresponds to a tilt angle of 0$^{\circ}$. For Na$^{+}$ and Mg$^{2+}$, the tilt angle from MD is 38±24$^{\circ}$, and 15±12$^{\circ}$, respectively, see Fig. \ref{fig:Orientation from MD}. Note that ±24$^{\circ}$ for Na and ±12$^{\circ}$ for Mg is not an uncertainty of the MD simulations, but a measure of the orientational disorder in the first solvation shell. 

To explore the effects of disorder on the spectra, we have calculated IRD spectra for for Na$^{+}$[H$_{2}$O]$_{6}$ and Mg$^{2+}$[H$_{2}$O]$_{6}$ in the HF approximation. IRD spectra as function of tilt are shown in Fig. \ref{fig:orientation}.

Starting with the case of ideal orientation, $\alpha$ = 0$^{\circ}$, the Na and Mg IRD spectra are simpler in the sense that they contain fewer components. This is due to the higher symmetry of the case of ideal orientation as compared to the other cases. This warrants identifying the different spectral regions with holes in the \textit{1b$_{1}$}, \textit{3a$_{1}$}, and \textit{1b$_{2}$} molecular orbitals. The meaning of this assignment is that the cluster orbitals can be seen as mainly linear combinations of specific orbitals of the individual water molecules. For both Na and Mg,  \textit{3a$_{1}$} dominates strongly, while \textit{1b$_{1}$} has no appreciable intensity. The \textit{1b$_{2}$} peak is weak for Mg, and practically absent for Na. 

Turning to the case of tilting, we see that both the Na and Mg IRD spectra consist of more energetically spread components, shifted towards higher energy.  This could be described as a relative weakening of the \textit{3a$_{1}$} and strengthening of the \textit{1b$_{1}$}, but this is complicated by the spectral overlap of states with main \textit{3a$_{1}$} and \textit{1b$_{1}$} character. Experimentally, the Na spectrum exhibits intensity in the \textit{1b$_{1}$} and\textit{ 3a$_{1}$} regions of approximately equal intensity, whereas for Mg the spectrum is dominated by \textit{3a$_{1}$}, with successively less intensity in the \textit{1b$_{1}$ }and \textit{1b$_{2}$ }regions. Comparing the calculated and experimental spectra, this indicates that the molecules on average are closer to the ideal orientation for Mg$^{2+}$ than for Na$^{+}$. Our MD simulations, see Fig. \ref{fig:Orientation from MD},  indicate that for Mg$^{2+}$[H$_{2}$O]$_{6}$, the tilt angle $\alpha$ is $\sim$15±12$^{\circ}$, while for Na$^{+}$[H$_{2}$O]$_{6}$  $\alpha$ is $\sim$38±24$^{\circ}$. This reflects the increase of ion-water interaction with increasing ion charge, causing the solvation shell to be more ordered around Mg$^{2+}$ than around Na$^{+}$. The  experimental spectra will consequently reflect a wider spread of orientations for Na$^{+}$ than for Mg$^{2+}$. The calculated IRD spectra for Mg at $\alpha$ = 10$^{\circ}$ and 20$^{\circ}$ agree quite well with the experimental one with a sharp dominating \textit{3a$_{1}$} peak. For Na at $\alpha$ = 40$^{\circ}$, the calculations predict substantial intensity in both the  \textit{3a$_{1}$} and the \textit{1b$_{1}$} regions, in reasonable agreement with the experimental spectrum. 

\begin{figure}[!htbp]
    \centering
    \includegraphics[width=0.75\linewidth]{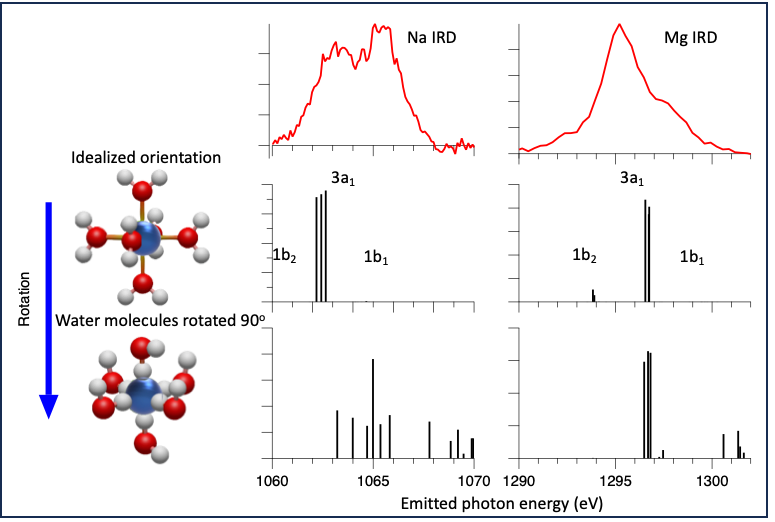}
    \caption{(Top panels) Experimental IRD spectra compared to (middle and bottom panels) calculated IRD spectra for Na$^{+}$[H$_{2}$O]$_{6}$ and Mg$^{2+}$[H$_{2}$O]$_{6}$ for both the ideal rotation of the H$_2$O molecules where the oxygen atom points towards the metal center, as well the H$_2$O molecules rotated so that a hydrogen atom points towards the metal center, compare to Fig. \ref{fig:orientation}. The same energetic shift as in Fig. \ref{img:spectra} in the main paper has been applied to the theoretical spectra.}
    \label{fig:rotation}
\end{figure}

In addition to tilting, disorder can be in the form of a rotation so that one of the hydrogen atoms faces towards the ion, see Fig. \ref{fig:rotation}. In the idealized orientation, the spectral shape for both Na and Mg is strongly dominated by the \textit{3a$_1$} peak. For Na, rotation makes the spectral intensity spread out in the region of the \textit{1b$_1$} peak. For Mg, the effect is similar, but less pronounced. Disorder in the form of  tilt and rotation thus have similar effects in the spectral shape. 

As noted in the main text, while the experimental and calculated Mg IRD spectra agree well, there is a discrepancy for Na in terms of relative intensities of \textit{1b$_{1}$$_{}$} and\textit{ 3a$_{1}$}. Experimentally \textit{1b$_{1}$} is somewhat stronger than \textit{3a$_{1}$}, whereas \textit{3a$_{1}$} is stronger in the calculated spectra. Given the dependence of the relative intensities on the molecular orientation, this discrepancy could be due to the disorder for Na$^{+}$[H$_{2}$O]$_{6}$ being underestimated in the calculations using the representative structure.

A complete simulation of the experimental spectra should include a sum of spectra calculated for a statistically representative sampling of geometries obtained as snapshots from MD simulations, which is beyond the scope of this paper. While our simulations for some selected deviations from ideal orientation are not exhaustive, they clearly demonstrate the sensitivity of IRD to the degree of orientational (dis)-order in the solvation shell.

\subsection*{IRD and Ion Pairing}
\begin{figure}[!htbp]
    \centering
    \includegraphics[width=0.8\linewidth]{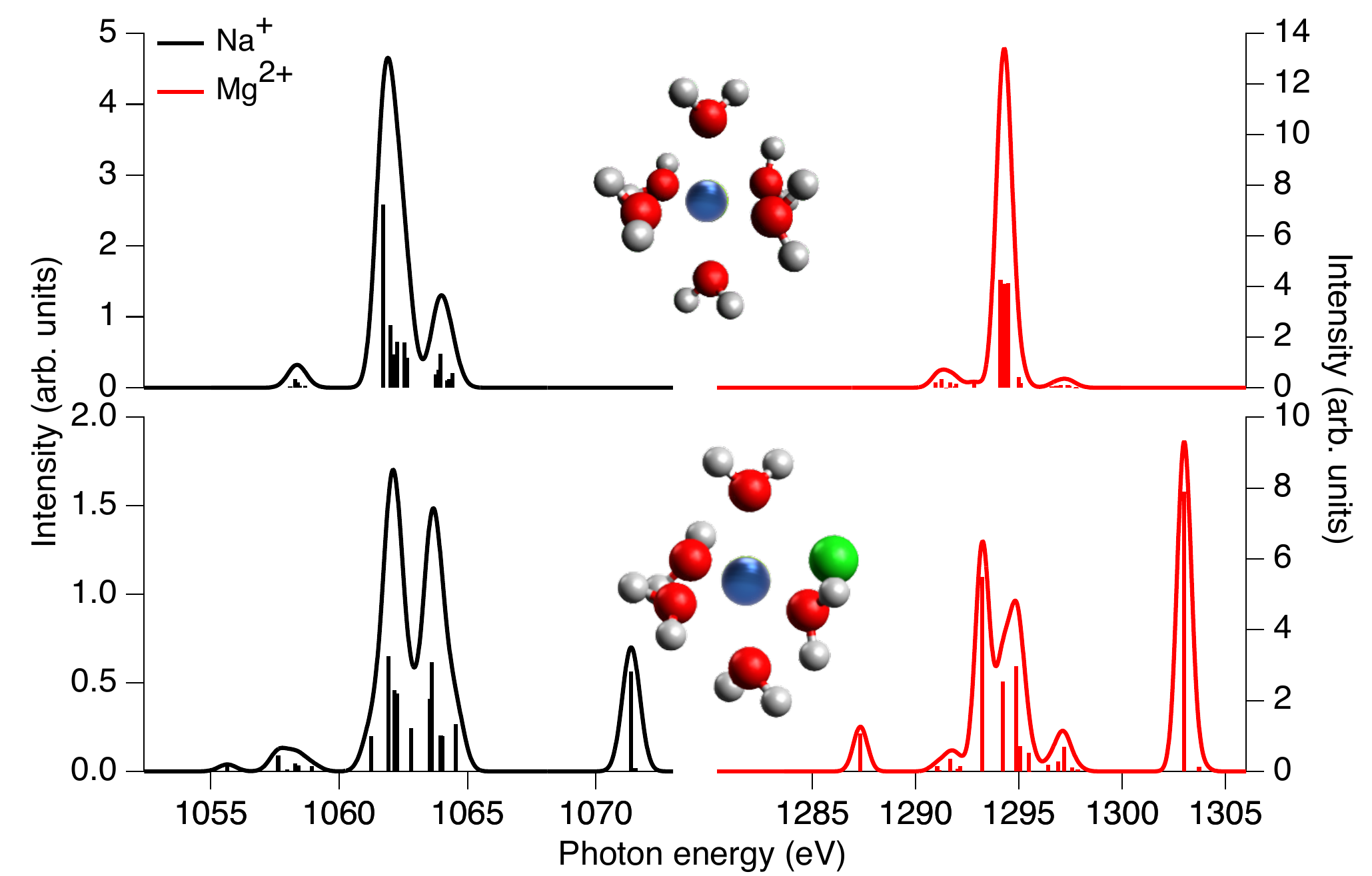}
    \caption{XES spectra computed by DFT and ground-state approximation for a metal cation (blue), either Na$^{+}$ (black lines) or  Mg$^{2+}$ (red lines), surrounded by six water molecules, i.e., completely hydrated (top panel), and five water molecules and a Cl$^{-}$ ion (green), i.e., a contact ion pair (bottom panel).}
    \label{fig:ionpairing}
\end{figure}

At low concentrations, the metal cations and their Cl$^-$ counter ions are both surrounded by solvation shells consisting of only water. With increasing concentration, there may also be ion pairing in the form of a Cl$^{-}$ counter ion replacing a water molecule in the solvation shell of the metal cation. Based on 
a combination of Raman spectroscopy, MD, and quantum chemical calculations, the existence of such Na$^+$ - Cl$^-$ contact ion pairs with a frequency increasing with concentration has been reported for NaCl in water \cite{Wang2023}. To investigate the effects of ion pairing on IRD, Fig. \ref{fig:ionpairing} compares the computed XES spectra for M$^{q}$[H$_{2}$O]$_{6}$ and M$^{q}$[H$_{2}$O]$_{5}$Cl$^{-}$ containing a contact ion pair, for M$^{q}$ = Na$^{+}$ and Mg$^{2+}$.

The main predicted  change is the appearance of a new band for M$^{q+}$[H$_{2}$O]$_{5}$Cl$^{-}$ at  $\sim$8 eV higher energy than the IRD feature of M$^{q+}$(H$_{2}$O)$_{6}$. A simple analysis of the valence molecular orbital involved in this transition shows that it is mainly due to a small mixing of the Cl \textit{3p} orbital with the M $2p$ orbital. We do not observe any such transitions in the experimental spectra, indicating that the fraction of ion pairing is low in the measured sample, in agreement with the results from Ref. \cite{ICDmanuscript}. The simulations, however, clearly show that IRD would be sensitive to such ion pairing. We conclude that IRD  is able to probe the composition of the solvation shells, which in the investigated cases are strongly dominated by water.

\newpage


\end{document}